\documentclass{article}

\usepackage{amsmath,amsfonts,amsbsy,amssymb,fullpage,makeidx,array,graphicx}
\usepackage{microtype}
\usepackage{amsmath}
\usepackage{amssymb}
\usepackage{mathtools}
\usepackage{rotating}
\usepackage{epstopdf}
\usepackage{caption}
\usepackage{lipsum}
\usepackage{afterpage}
\usepackage{subfigure}
\usepackage{bm}
\usepackage[normalem]{ulem}
\usepackage[nomarkers,nolists]{endfloat}

\def\meanIapp{\langle I_{app} \rangle}

\def\meangsyn{\langle g_{syn} \rangle}

\def\meanWj{\langle W_{jump} \rangle}
\def\meanRb{\langle R_i|\beta \rangle}

\begin{document}

\author{Wilten Nicola and Sue Ann Campbell\\ 
{\small Department of Applied Mathematics, University of Waterloo}\\{\small Waterloo ON N2L 3G1}}
\title{Mean-Field Models for Heterogeneous Networks of Two-Dimensional Integrate and Fire Neurons}

\maketitle

\begin{abstract}
We analytically derive mean-field models for all-to-all coupled networks of heterogeneous, adapting, two-dimensional integrate and fire neurons.  The class of models we consider includes the Izhikevich, adaptive exponential, and quartic integrate and fire models.   The heterogeneity in the parameters leads to different moment closure assumptions that can be made in the derivation of the mean-field model from the population density equation for the large network.  Three different moment closure assumptions lead to three different mean-field systems.  These systems can be used for distinct purposes such as bifurcation analysis of the large networks, prediction of steady state firing rate distributions, parameter estimation for actual neurons, and faster exploration of the parameter space.  We use the mean-field systems to analyze adaptation induced bursting under realistic sources of heterogeneity in multiple parameters. Our analysis demonstrates that the presence
of heterogeneity causes the Hopf bifurcation associated with the emergence of bursting to change from sub-critical to super-critical. This is confirmed with numerical simulations of the full network
for biologically reasonable parameter values.  This change decreases the plausibility of adaptation being the cause of bursting in hippocampal area CA3, an area with a sizable population of heavily coupled, strongly adapting neurons.     
\end{abstract}

\section{Introduction}

As computers become more powerful, there is a move to numerically simulate 
larger and larger model networks of neurons \cite{Izbig}. While simulation is
useful for confirming observed behavior it is not as helpful
in determining the mechanisms underlying the behavior. The tools of dynamical
systems theory, such as bifurcation analysis can be useful in this regard
when studying single neuron or small network models. However,
they are not viable for large networks, especially if the neurons are not 
identical. Thus, a common approach is to try to extrapolate large network
behavior from detailed analysis of the behavior of individual cells 
or small networks \cite{SBC}. This can be problematic as networks can have
behavior that is not present in individual cells. For example, individual
neurons that are only capable of tonic firing when isolated may burst when 
coupled in a network \cite{VreeswijkandHansel}.  Further, large networks
may exhibit behavior not present in smaller networks. For example, 
Dur-e-Ahmad et al. studied bursting in networks ranging in size from 
2 cells to 100. They found that bursting occurred in a larger range of 
parameters for larger networks \cite[Figure 7]{us}.

Given the role of bursts in networks of neurons, it is important to understand how a network transitions (bifurcates) from a non-bursting behavior, to bursting.   Bursting has been suggested to be a fairly important and information dense firing mode for neurons.   For example, single bursts can induce long term potentiation and depression in the hippocampus, which are important proccesses for learning and memory \cite{Lisman}.  Additionally, bursts have been found to carry more information about an animals position in space then isolated spikes alone as place fields have been found to be more accurately defined when considering bursts alone \cite{Lisman}.  Additionally, the mechanism we analyze in this paper, adaptation induced bursting has also been suggested as a biologically plausible mechanism for the generation of grid cells via oscillatory interference of the bursting \cite{Hasselmo}.  However, adaptation induced bursting is a network level phenomenon and we cannot apply bifurcation analysis directly to a network of adapting neurons.  

Mean-field theory offers one approach to bridge this gap. In applying this
theory, one usually derives (or suggests) a low dimensional system of differential equations that govern the moments of different variables across the network  \cite{Bressloff}.   
For example, for a network of all-to-all coupled Izhikevich neurons \cite{Izhikevich}, one can derive a two dimensional system of differential equations 
for the mean adaptation variable, and the mean synaptic gating variable \cite{us2}.  
Mean-field systems enable one to conduct network level bifurcation analysis and
hence to test different hypotheses about large network behavior.  For 
example, Hemond et al. \cite{Hemond} found that when uncoupled, the majority 
of hippocampal pyramidal neurons in region CA3 do not display bursting.  
This is contradictory to the observation that burst firing is ubiquitous 
this region \cite[Section 5.3.5]{hippobook}.  A possible explanation for 
these contradictory observations is that bursting is a network level 
phenomenon.  This hypothesis has been tested
using bifurcation analysis of the mean-field system derived from a network 
of Izhikevich neurons \cite{us}. In particular, it was shown that for a 
network of identical all-to-all coupled neurons fit to experimental data 
from \cite{Hemond} for CA3 pyramidal cells, bursting occurs for a large
range of synaptic conductances and applied currents if the spike frequency
adaptation in the neurons is sufficiently strong. 

Hemond et al.~ \cite{Hemond} also observed that the pyramidal 
neurons in their study were heterogeneous, in particular, the neurons
had different degrees of spike frequency adaptation.  When the study of 
\cite{us} was extended to a network of two 
homogeneous subpopulations of hippocampal neurons with different degrees of 
spike frequency adaptation, the mean-field equations predict, and numerical 
simulations confirm, that the region in the parameter space where bursting 
occurs decreases in size \cite{us2}.  This would seem to indicate that 
adaptation induced bursting may not be robust to heterogeneity, however, 
it is unknown how robust this network level bursting is to heterogeneity in 
different parameters.  An extension of the mean-field system in \cite{us2} 
to large networks of heterogeneous neurons, is needed to fully analyze the 
robustness of adaptation induced bursting.
 
The application of mean-field analysis to large networks of heterogeneous 
neurons has been limited to networks of one dimensional integrate and fire 
neurons with heterogeneity in the applied current\cite{HM1,HM,Vlad}. 
Hansel and Mato \cite{HM1,HM} analyze a network of all-to-all coupled
quadratic integrate and fire neurons consisting of two 
subpopulations: one excitatory and one inhibitory. They showed analytically
that the tonic firing asynchronous state can lose stability either to a
synchronous tonic firing state or a bursting state.
Vladimirski et al. \cite{Vlad} analyze a network of all-to-all coupled 
linear integrate and fire neurons subject to
synaptic depression.  This model was used to study network induced bursting 
in the developing chick spinal cord.  Their derivation of the mean-field 
model is based on temporal averaging of the fast synaptic gating variable, 
in addition to the usual network averaging. This results in a model that only 
involves the distribution of slow synaptic depression variable and the 
distribution of firing rates.  Vladimirski et al. note that, for their model, 
increased heterogeneity tends to make population induced bursting more robust. 
They also note that one cannot understand the behaviour of their network 
with a single slow, network averaged synaptic depression variable.

The motivation for the present paper is to explore the effect of heterogeneity 
in parameters on network induced bursting when adaptation is the primary 
negative feedback acting on the individual firing rates.  To this end,
we introduce a set of mean-field equations for networks of heterogeneous 
two-dimensional integrate and fire neurons.  While the specific neural model 
we consider is for neurons with adaptation, our derivation is quite general
and could be applied to other integrate and fire models.
In contrast with \cite{Vlad} our derivation of the mean-field model does not 
use temporal averaging thus we end up with a mean-field system which involves 
the network averaged
synaptic gating variables as well as the distribution of adaptation variables. 
We also allow for heterogeneity in more than one parameter.
Our approach is a generalization of that used for homogeneous
networks of two dimensional integrate and fire neurons \cite{us2}, however, 
in the heterogeneous case it turns out that 
there are actually multiple mean-field models, as different assumptions can 
be made 
during the derivation.  This leads us to three distinct mean-field systems, each derived under different assumptions, and used for different purposes.  Together, these sets of equations allow us to do bifurcation analysis on large networks, as in the homogeneous case. We show that the bifurcation structure of the heterogeneous network differs both qualitatively and quantitatively from the homogeneous network case.  We discuss the implications of this for network induced bursting in the hippocampus.    

 When considering a homogeneous network, the mean-field variables are a good approximation for the variables of every neuron.  However, this is not the case for a heterogeneous network.  If the heterogeneity is large, then the neuron variables may be also widely distributed, rendering information about the first moments less useful.   This also implies that the behavior of any individual neuron is less predictable with a mean-field system than it was in the homogeneous case. One of our mean-field systems addresses this problem, giving
information about the distributions of the variables instead of just the mean. 

When considering a model for a specific heterogeneous network of neurons, one stumbling block 
is determining the distribution of parameters.  Estimates of the distribution can be made through direct intracellular recording of a sufficient number of neurons, and conventional measurements of the biophysical properties (membrane capacitance, voltage threshold, etc.).  Unfortunately, this is a very time consuming and intensive process.  What is needed is a way of measuring the biophysical parameters of multiple neurons simultaneously.   There are a few ways to sample multiple neurons such as multi-unit recordings using tetrode arrays, or two-photon microscopy techniques \cite{Buzsaki04,Grewe}. However, these techniques typically only tell us about spike times of large (dozens to hundreds of neurons) networks \cite{Buzsaki04}.  While an impressive accomplishment, this still does not tell us anything directly about the biophysical properties of the neurons that caused those spikes.  
In this paper we use a mean-field system to determine an approximate distribution of firing rates for a network given a known distribution of parameters.  This is an unusual state of affairs for a mean-field system, as these kinds of systems seldom give information about entire distributions.   More importantly however, using a mean-field system, we can invert a distribution of steady state firing rates (which can be obtained from multi-unit recordings)  to obtain a distribution of parameters.  In fact, this can be done at the individual neuron level, to determine the parameter value for any particular neuron.  This allows one to estimate different biophysical parameters, which are difficult to measure at the network level, using easy to measure firing rate distributions.  However, the assumptions required for the numerical accuracy of the estimation are fairly strong.   

The plan for our article is as follows.
Section 1.1 introduces the general class of adapting two-dimensional integrate 
and fire neurons used in our network models.  This class was introduced by 
Touboul \cite{Touboul2008}, who also completed the bifurcation analysis of the 
single, uncoupled neuron.   Population density methods are briefly introduced 
in section 1.2, as a population density equation serves as an intermediate 
step to obtain our mean-field models.   Section 2 begins with a review of
mean-field theory and the equations for the homogeneous network. This is 
followed, in sections 2.1-2.3, by the derivation of the three mean-field 
systems for the heterogeneous network.   A comparison of numerical 
simulations of these mean-field systems and the full network is the subject
of section 2.4.  Applications of mean-field theory to networks with a 
single heterogeneous parameter can be found in Section 3 including bifurcation 
analysis (Section 3.1), distributions of parameters and firing rates (Section 3.2) and using mean-field theory for parameter estimation from firing rate data (Section 3.3).  Applications of mean-field theory to networks with multiple sources of heterogeneity are included in section 4.  
A discussion of our work and its implications can be found in section 5.

\section{Materials and Methods} 
\subsection{Nonlinear Integrate and Fire Neurons}
We consider a network of two-dimensional integrate and fire models of the form 
\begin{eqnarray}
\dot{v} &=& F(v) - w + I \label{de1} \\ 
\dot{w} &=& a(bv - w) \label{de2},
\end{eqnarray}
where $v$ represents the nondimensionalized membrane potential, and $w$ serves as an adaptation variable.  Time has also been non-dimensionalized.  
The dynamical equations (\ref{de1})-(\ref{de2}) are supplemented by the following discontinuities 
\begin{equation}
v(t_{spike}^-) = v_{peak} \quad \Rightarrow \quad 
\begin{array}{rcl}
v(t_{spike}^+) &=& v_{reset},\\
w(t_{spike}^+) &=& w(t_{spike}^-) + w_{jump}.
\end{array} \label{eq_reset}
\end{equation}
This particular notation was formally introduced by Touboul \cite{Touboul2008}, along with a full bifurcation analysis of this general family of adapting integrate and fire neurons.  Members of this family include the Izhikevich model  \cite{Izhikevich}, the adaptive exponential (AdEx) model \cite{AdEx,Ad2}, and Touboul's own quartic model \cite{Touboul2008}.    

The methods of this paper can be applied to a network of any particular neuron belonging to 
this general family, and thus all derivations are done for this model. For the numerical 
examples, however, we only consider the Izhikevich neuron:
\begin{eqnarray}
C\dot{V}_i &=& k(V_i-V_T)(V-V_R) - W_i + I_{app,i} \label{IzV}\\
\dot{W}_i &=& \frac{\eta(V_i-V_R) - W_i}{\tau_W}\\
V_i(t_{spike}^-) &=& V_{peak}\quad \Rightarrow \quad
\begin{array}{rcl}
V_i(t_{spike}^+) &=& V_{reset} \\
W_i(t_{spike}^+) &=& W_i(t_{spike}^-) + W_{jump,i}, \label{Izhom_jump}
\end{array}
\end{eqnarray}
In dimensionless form, this is given by \eqref{de1}-\eqref{eq_reset} with $F(v)=v(v-\alpha)$ in addition to dimensionless versions of the discontinuities \eqref{Izhom_jump}.
The application to other neural models is straight forward, see \cite{us2} where the 
homogeneous mean-field theory has been derived and tested for both the AdEx and the Izhikevich 
models. 

Networks of these neurons can be coupled together through changes in the synaptic conductance.   
The synaptic conductance of post-synaptic neuron $i$ due to presynaptic neurons $j=1,2,\ldots,N$
is given by 
\begin{equation}
g_{i}(t) = g_is_i(t) =\frac{{g_i}}{N} \sum_{j=1}^N s_{ij}(t),\label{si_def}
\end{equation}
where ${g}_{i}$ denotes the maximal synaptic conductance of neuron $i$ 
and $s_i(t)$ denotes the total proportion of postsynaptic ion channels open 
in the membrane of neuron $i$.  The time dependent variable 
$s_{ij}(t)$ represents the proportion of postsynaptic ion channels open in 
the membrane of neuron $i$ as a result of the firing in neuron $j$.  

The changes in $s_{ij}(t)$ that occur after a spike are often modeled as transient pulses. 
For example, if neuron $j$ fires its $k$th action potential at time $t=t_{j,k}$, then the variable $s_{ij}(t)$ at time $t$ is given by 
\begin{equation}
s_{ij}(t) = \sum_{t_{j,k}<t}E(t-t_{j,k}).
\end{equation}
There are different functions proposed for $E(t)$ in the literature including the simple exponential, the alpha synapse, and the double exponential.  We primarily consider the simple exponential synapse 
\begin{equation}
E(t)  = {s_{jump}}\exp{ \left(\frac{-t}{\tau_s}\right)}, 
\end{equation}
which is governed by the ordinary differential equation 
\begin{equation}
\frac{ds_{ij}(t)}{dt} = -\frac{s_{ij}}{\tau_s} + s_{jump} \sum_{t_{j,k}<t} \delta(t-t_{j,k}).
\label{sij_eq}
\end{equation}

In the rest of the paper, we assume all-to-all connectivity and that 
the synaptic parameters $s_{jump}$, and $\tau_s$ are the same for every
synapse. In this case we may set $s_i(t) = s(t)$ for all $i$, as each 
postsynaptic neuron receives the same summed input from all the presynaptic 
neurons.   Then, using \eqref{si_def} and
\eqref{sij_eq}, the network of all-to-all coupled neurons that we consider is given by the following system of discontinuous ODE's:
\begin{eqnarray}
\dot{v}_i &=& F(v_i) - w_i + I_i + g_is(t)(E_r - v_i),\\
\dot{w}_i &=& a_i(bv_i - w_i),\\
\dot{s} &=& -\frac{s}{\tau_s} + \frac{s_{jump}}{N}  \sum_{j=1}^N\sum_{t_{j,k}<t} \delta(t-t_{j,k}),\\
\end{eqnarray}
\begin{equation}
v_i(t_{j,k}^-) = v_{peak} \quad \Rightarrow \quad 
\begin{array}{rcl}
v_i(t_{j,k}^+) &=& v_{reset},\\
w_i(t_{j,k}^+) &=& w_i(t_{j,k}^-) + w_{jump}.
\end{array} 
\end{equation}
for $i=1,2,\ldots N$.  

In the examples, we consider one or more parameters as the sources of 
heterogeneity. 
However, to simplify the notation 
in the derivations, we use the vector $\bm\beta$ to represent all
the heterogeneous parameters. Then, denoting the state variables $v$ and $w$ 
as the vector $\bm{x}$,   we can write the equations for the individual 
oscillator as 

\begin{equation}
\dot{\bf x}={\bf G}({\bf x},{\bm\beta},s)=\left(\begin{array}{l}
 G_1(\bm{x},\bm{\beta},s)\\
 G_2(\bm{x},\bm{\beta})
\end{array}\right)\end{equation}

Given a specific heterogeneous parameter, $G_1$ and $G_2$ may not 
depend on $\bm\beta$, or all of the components of $\bm\beta$.  However, 
for the sake of simplicity, we include the dependence in both equations. 

Our numerical examples are restricted to the Izhikevich neural model, 
and we primarily consider the driving current $I_i$ of each neuron, the 
synaptic conductance $g_i$ and the adaptation jump size, $w_{jump}$
as the source of heterogeneity.  However, the mean-field equations we 
derive can be applied to any of the two-dimensional adapting integrate 
and fire models, with any heterogeneous parameter or set of parameters. 

Finally, we note that in many applications, $b$ is a small parameter, and 
thus the $bv$ term can be dropped in $G_2$.  We do this in all our numerical 
studies.  However, one can can still derived appropriate mean-field 
equations if this term is present (see discussion in \cite{us2}), 
and thus we have left the term in the derivations. 

\subsection{The Population Density Equation}
The population density function, $\rho(\bm{x},t)$ determines the density of neurons at a point in phase space, $\bm{x}$, at time $t$.  Consider first the
case of a homogeneous network, i.e., all the oscillators have the same
parameter values, denoted by $\bm\beta$.  In the limit as $N\rightarrow \infty$, 
one can derive the following evolution equation for 
the population density function:
\begin{eqnarray}
\frac{\partial \rho(\bm{x},t)}{\partial t} = -\nabla \cdot \textbf{J}(\bm{x},\bm{\beta},s,t) \label{continuityequation}
\end{eqnarray}
where $\textbf{J}$ is given by
 \begin{equation}
 \textbf{J}(\bm{x},\bm{\beta},s,t) = \textbf{G}(\bm{x},\bm{\beta},s)\rho(\bm{x},t) = \left(J^V,J^W \right).
 \end{equation}
and must satisfy the boundary condition 
\begin{equation}
J^V(v_{peak},w,\bm\beta,s,t)=J^V(v_{reset},w+w_{jump},\bm\beta,s,t). \label{BC1}
\end{equation}
In the same limit, the differential equation for $s$ converges to 
 \begin{equation}
 s' = -\frac{s}{\tau_s} + s_{jump}\int_W J^V(v_{peak},w,s,\bm{\beta},t)\,dw
 \end{equation}
where the integral term is actually the network averaged firing rate, 
which we denote as $\langle R(t) \rangle$.  
Derivations of equation (\ref{continuityequation}) can be found in various sources \cite{NT,OKS}.  

Equation \eqref{continuityequation} is frequently referred to as the continuity equation and it has various applications besides its use as an intermediate step in mean-field reductions.  For example, the equation has been used to determine the stability of the asynchronous tonic firing state by various authors \cite{AbbottandVreeswijk,HM,Sirovich,Strog1,Vrees,VAE}.  These papers predominantly 
consider homogeneous networks of linear integrate and fire neurons. The
exception is the work of Hansel and Mato \cite{HM} which considers heterogeneity
in the applied current. One can study stability of various firing states
using spectral analysis or other analytical treatments of this equation 
\cite{AbbottandVreeswijk,HM,Knight2,Knight,Sirovich,Strog1,Vrees}. However, these
approaches are too complicated for the models we consider in this paper.  

Now consider a heterogeneous network where the parameters vary from oscillator 
to oscillator, but are static in time. Then one can rewrite the equations for 
the individual oscillator as 
   \begin{eqnarray}
\dot{v}_i &=&  G_1(\bm{x}_i,\bm{\beta}_i,s),\\
\dot{w}_i &=& G_2(\bm{x}_i,\bm{\beta}_i), \\
\dot{\bm\beta}_i &=& 0. 
\end{eqnarray}
 In this case the flux contribution due to $\bm\beta$ is 0, and the evolution equation for the network is given by 
\begin{eqnarray}
\frac{\partial \rho(\bm{x},\bm\beta,t)}{\partial t} = -\nabla \cdot \textbf{J}(\bm{x},\bm{\beta},s,t) 
\label{ceq_het}
\end{eqnarray}
The density now has the vector of parameters, $\bm\beta$, as an independent 
variable. The flux consists of the vector $(J^V,J^W,0)$, with $\bm\beta$ as an independent variable, as opposed to a fixed constant.  If the parameters are time varying however, the final component of the flux will be non-zero.   The equation for $s$ is also different in the heterogeneous case: 
 \begin{equation}
 s' = -\frac{s}{\tau_s} + s_{jump}\int_W \int_{\bm\beta}J^V(v_{peak},w,s,\bm{\beta}',t)\,dw\,d\bm\beta'.
 \end{equation}

While the evolution equation \eqref{ceq_het} is an exact representation for the network in the large network limit, it is difficult to work with analytically.  Additionally, as the dimensions of the PDE become large, it becomes difficult to find numerical solutions efficiently \cite{LT}.  However, mean-field reductions of the network can be used to reduce the population density PDE to a system of nonlinear switching ODEs that governs the moments of the distribution.    Unlike the PDE, the system of ODEs is tractable using bifurcation theory, at least numerically.   Furthermore, we show that in the heterogeneous case, the resulting mean-field systems can yield more information than just the type of bifurcation that the network can undergo.

\subsection{Mean-Field Theory}\label{MFTsec}
In the homogeneous case, the mean-field system of equations for an all-to-all coupled Izhikevich network was derived in \cite{us2}.  We present here a quick summary of this derivation.  In order to derive a mean-field system of equations, one first needs to reduce the PDE for $\rho(\bf x ,t) = \rho(v,w,t)$ by a dimension.  This is done by first writing the density in its conditional form:
\begin{equation}
\rho(v,w,t) = \rho_V(v,t) \rho_W(w|v,t) 
\end{equation}
and then integrating the continuity equation with respect to $w$.  This yields the one dimensional PDE
\begin{equation}
\frac{\partial \rho_V(v,t)}{\partial t} = -\frac{\partial G_1(v,s,\langle w | v \rangle)\rho_V(v,t)}{\partial v}  = -\frac{\partial J(v,\langle w | v \rangle, s,t)}{\partial v},
\end{equation}
where the flux has been redefined to 
\begin{equation}
 J(v,\langle w | v \rangle, s,t) = \int_W J^V(v,w,s,t)\,dw .
\end{equation}
One can now make a first order moment closure assumption, $\langle w | v \rangle = \langle w \rangle$, and derive an approximate ODE for $\langle w \rangle$, which yields the system 
\begin{eqnarray*}
\frac{\partial}{\partial t}\rho(v,t) &=& - \frac{\partial }{\partial v}\left((F(v)- \langle w \rangle+ I + g(E_r-v)s)\left(\rho(v,t)\right) \right)\\
\langle w \rangle' &=&\ \frac{b\langle v\rangle-\langle w \rangle}{\tau_w} +  w_{jump}J(v_{peak},\langle w \rangle,s,t)\\
s' &=& - \frac{s}{\tau_s} + s_{jump}J(v_{peak},\langle w \rangle,s,t)
\end{eqnarray*} 
where the subscript on the density function has been dropped for convenience.   The details, and validity of the first order moment closure assumption that is used can be found in \cite{LT}.  We note however that the work in \cite{LT} was primarily with leaky integrate and fire neurons, as opposed to the two-dimensional adapting class we consider here.  However, it is a necessary assumption to proceed analytically.  
If we assume that the adaptation time constant, $\tau_w = \frac{1}{a}$ is large, one can apply a quasi-steady state approximation to derive a  system of switching ODE's for $\langle w \rangle$ and $s$:  
\begin{eqnarray}
\langle w\rangle'&=& \frac{b\langle v\rangle- \langle w\rangle}{\tau_w} + w_{jump}\langle R_i(t)\rangle \label{wone} \\
s'&=& -\frac{s}{\tau_s} + s_{jump}\langle R_i(t)\rangle  \label{stwo} \\
\langle R_i(t) \rangle &=&    \left\{     \begin{array}{lr}     \left[ \int_{V}\frac{dv}{F(v)-\langle w \rangle + I + g(e_r - v)s}  \right]^{-1} & : H(\langle w \rangle,s)\geq 0  \\      0 &:H(\langle w \rangle,s)<0   \end{array}   \right.\label{switch1}  
\end{eqnarray}
The switching manifold for the system, $H(\langle w \rangle,s)$ is given by: 
\begin{equation}   
H(\langle w \rangle,s) =I -\langle w \rangle  +  \min_{v}(F(v) + g(e_r - v)s).
\label{Hdef}
\end{equation}
Note that $H(\langle w \rangle,s)$ depends on the parameter(s) of the model, and thus for the heterogeneous case, we make this dependence explicit by writing $H(\langle w\rangle,s,\bm\beta)$.   As the computation for $\langle v \rangle$ is somewhat lengthy and is only outlined in the discussion of \cite{us2}, we have placed it in Appendix A.   

\begin{figure}
\centering
\subfigure[$I_{app}=4500$ pA, $g_{syn}=200$ nS]{\includegraphics[scale=0.5]{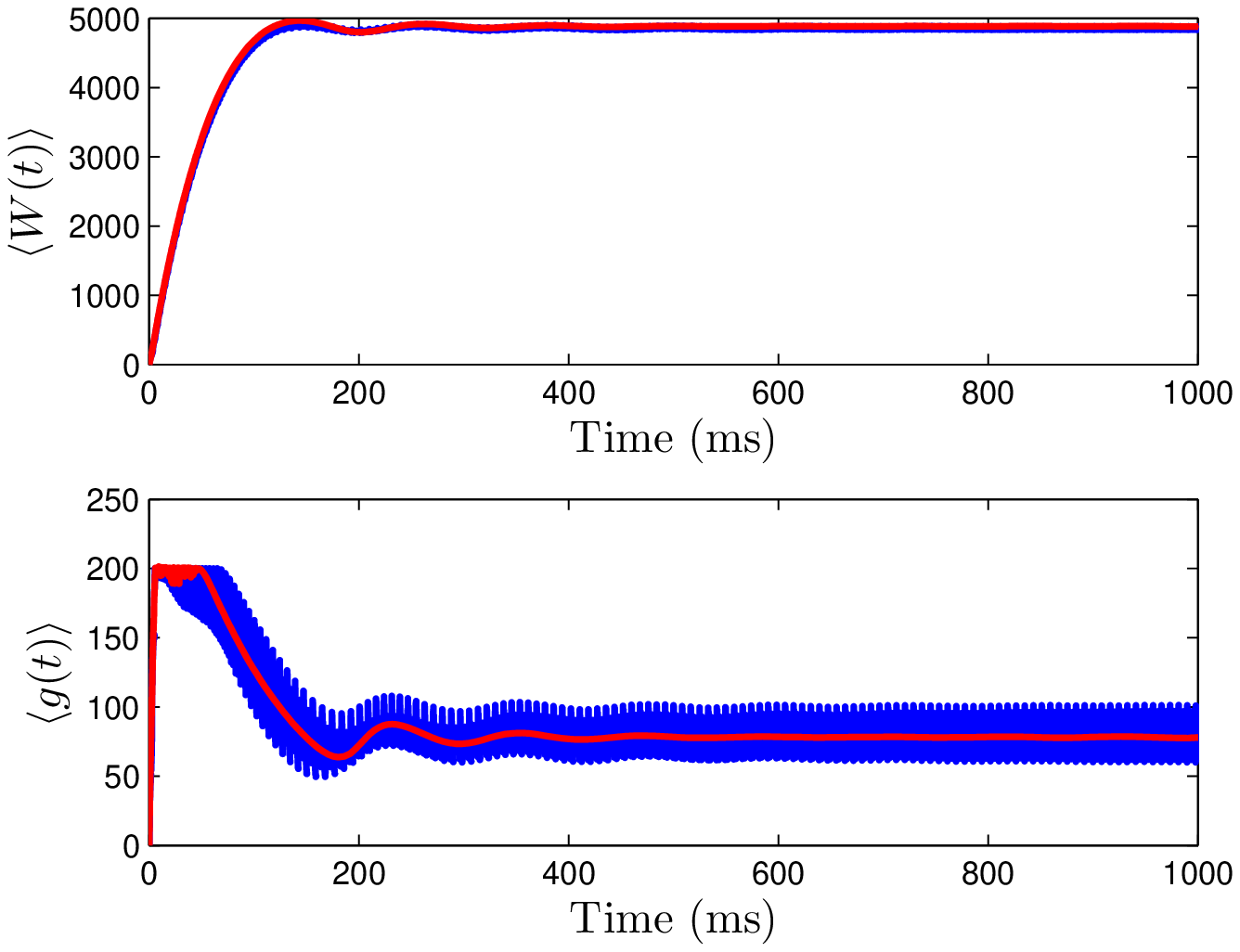}
\label{fig1a}}
\qquad 
\subfigure[Spike time raster plot for (a)]{\includegraphics[scale=0.5]{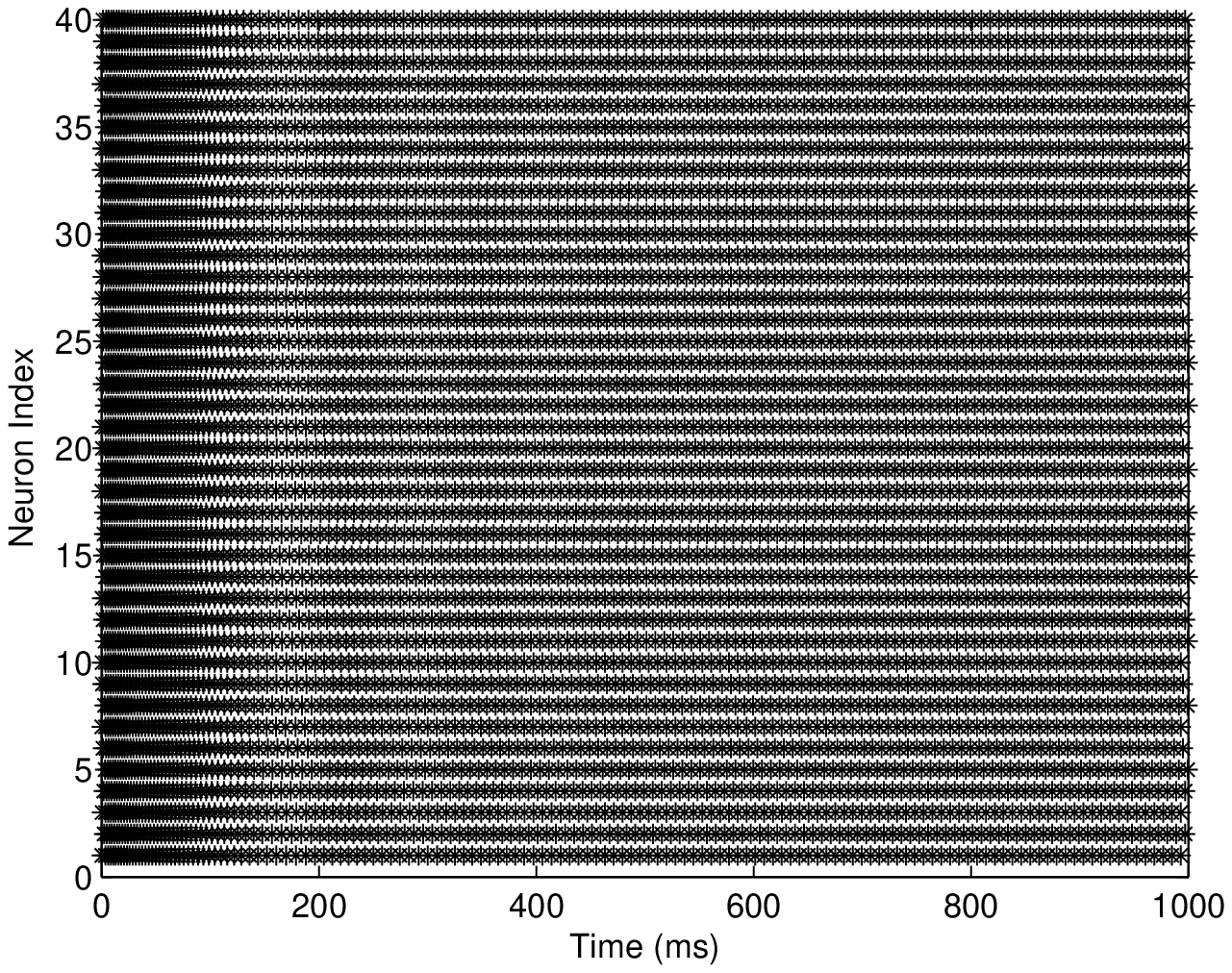}
\label{fig1b}}
\\ 
\subfigure[$I_{app}=3500$ pA, $g_{syn}=200$ nS]{\includegraphics[scale=0.5]{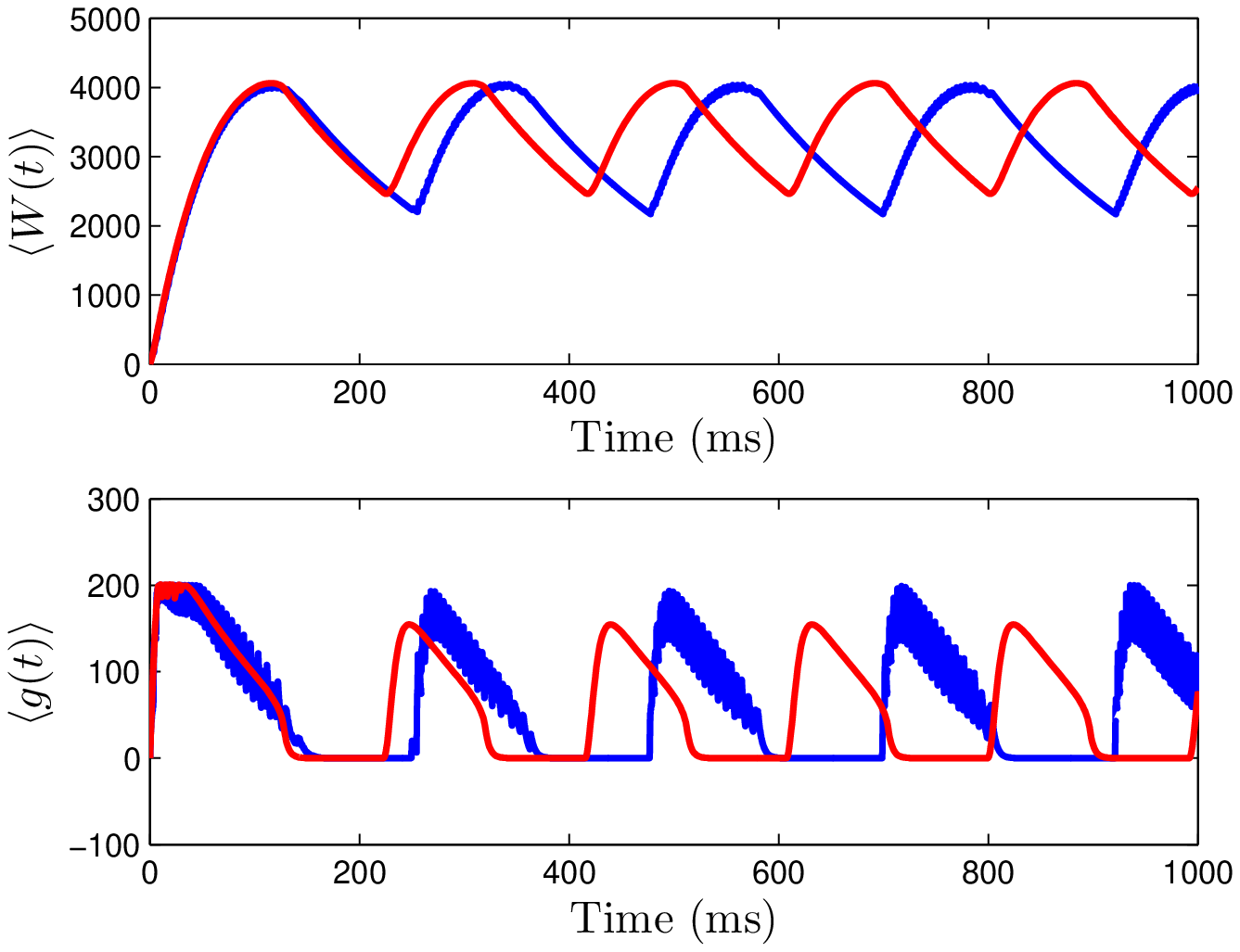}
\label{fig1c}}
\qquad 
\subfigure[Spike time raster plot for (c)]{\includegraphics[scale=0.5]{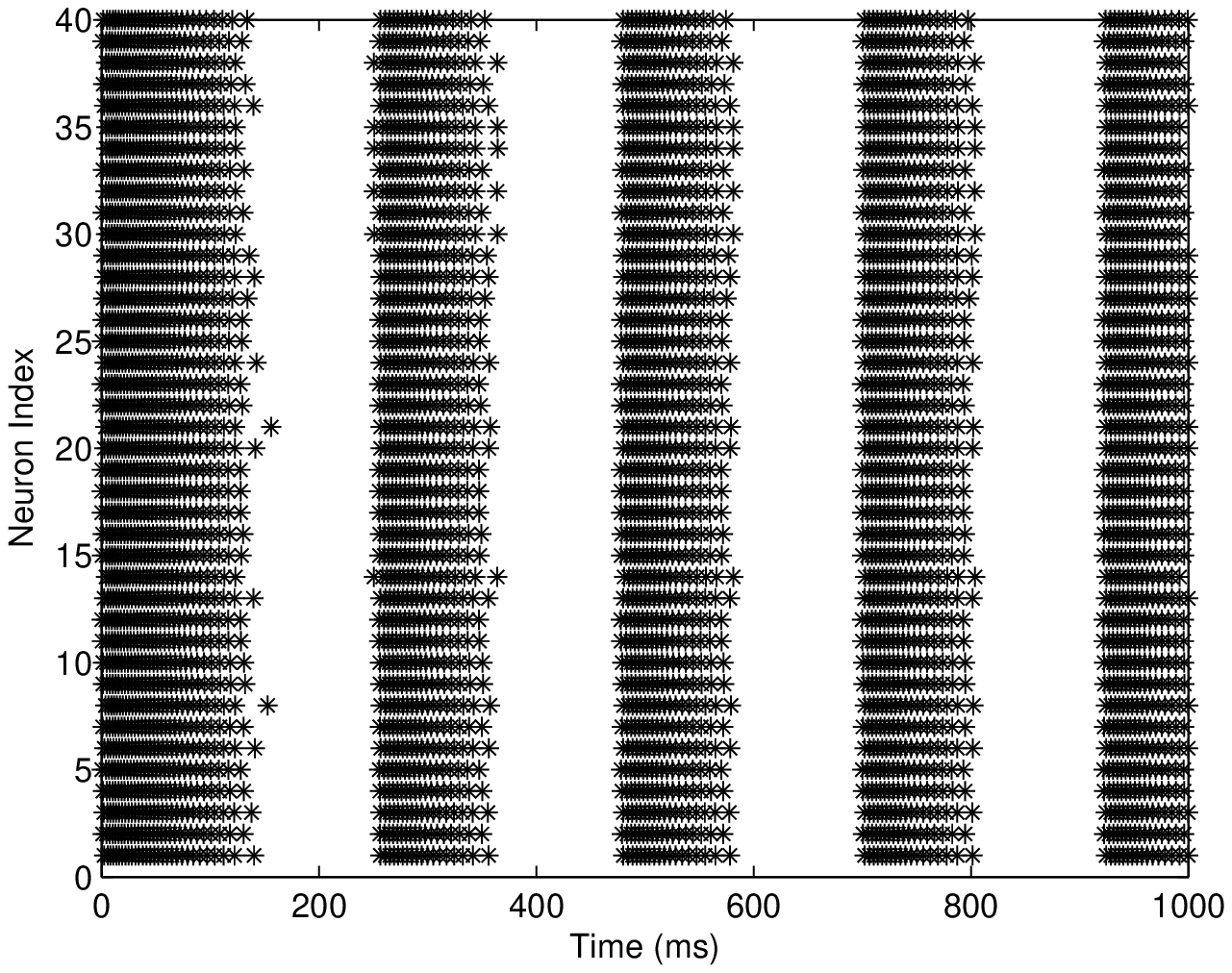}
\label{fig1d}}
        \caption{Numerical simulation of a homogeneous network of 1000 Izhikevich neurons, with parameters as shown.  The rest of the parameters can be found 
in Table 1.  Simulations of the mean-field equations (in red) and the mean values of the corresponding full network simulations (in blue) showing (a) tonic firing and (b) bursting.  (b),(d) Raster plots for 40 randomly selected neurons from the
network simulation in (a),(c).  The mean-field equations are fairly accurate both when the network is tonically firing and when it is bursting.} 
\label{fig1}
\end{figure}

For the Izhikevich neuron, equations \eqref{switch1}-\eqref{Hdef} become
\begin{eqnarray}
\langle R_i(t) \rangle &=&   \left\{ \begin{array}{lr} \left[
 \int_{V}\frac{dv}{v(v-\alpha)-\langle w \rangle + I + g(e_r - v)s}  \right]^{-1} & : H(\langle w \rangle,s)\geq 0  \\      
0 &:H(\langle w \rangle,s)<0 \end{array} \right.\label{Izswitch1}  \\
H(\langle w \rangle,s) &=& I - \langle w \rangle -\frac{(\alpha+gs)^2}{4} + ge_rs. \label{HIz} 
\end{eqnarray}
Note that in this case, we can evaluate $\langle R_i(t)\rangle$ explicitly: 
\[
\langle R_i(t) \rangle =  \left\{  \begin{array}{lr}     \frac{\sqrt{I - I^*(\langle w \rangle)}}{\arctan\left(\frac{ v_{peak}-\frac{\alpha+gs}{2}}{\sqrt{I-I^*(\langle w \rangle,s)}}\right)-\arctan\left(\frac{ v_{reset}-\frac{\alpha+gs}{2}}{\sqrt{I-I^*(\langle w \rangle,s)}}\right)}& : H(\langle w \rangle,s)\geq 0  \\      0 &:H(\langle w \rangle,s)<0   \end{array}   \right. 
\]
in addition to an approximation to $\langle v \rangle $
\begin{equation}
\langle v \rangle =  \left\{  \begin{array}{lr}  \frac{\langle R_i(t)\rangle}{2}\log\left(\frac{(v_{peak} - \frac{\alpha+gs}{2})^2 +H(\langle w\rangle,s)}{(v_{reset} - \frac{\alpha+gs}{2})^2 +H(\langle w\rangle,s)}\right) + \frac{\alpha+gs}{2}   & : H(\langle w \rangle,s)\geq 0  \\      \frac{\alpha+gs}{2} - \sqrt{-H(\langle w\rangle,s)}  &:H(\langle w \rangle,s)<0   \end{array}   \right.  \label{meanvhom}
\end{equation}
A comparison of solutions of these equations and the full network are shown in Figure \ref{fig1} for both the tonic firing and the bursting case.  

This system of equations is valid when $\tau_w \gg O(1)$, however the magnitude 
of $\tau_s$ is also significant.  While in the original derivation of 
\cite{us2}, $\tau_s = O(\tau_w)$ was suggested as a criterion for validity, this is not actually necessary.  One merely requires that $\tau_s$ not be significantly smaller then $O(1)$, the time scale of the PDE.  The reason for this is that if
the time scale of the ODE for $s$ is smaller than that of the PDE then the 
quasi-steady state approximation must be applied to the ODE for $s$ as well.
The requirements on the time constants are carried forward in the heterogeneous case.  

In our models, the timescale of the ODE for $s$ is typically between that 
of the PDE and that of the ODE for $w$, and we have not applied the
quasi-steady approximation to $s$. 
Applying a quasi-steady state approximation to both $s$ and the reduced PDE yields a more compact system, which is just an ODE for $\langle w\rangle$, however the analysis does not get any simpler.   The reason for this is two-fold: the ODE for $\langle w \rangle$ remains non-smooth, and the firing rate now has to be implicitly solved at each time step.  Thus, it is more convenient to apply the quasi-steady state approximation only to the partial differential equation.   

When parameter heterogeneity is added into the mix, it turns out that there are multiple ``mean-field" systems of equations that can be derived, by applying
different assumptions on the conditional moments. We outline three different assumptions that can be made, and the resulting system of mean-field equations in each case.

\subsubsection{Mean-Field System I}
We begin by writing out the density function in the conditional form
\begin{equation}
\rho({\bf x},\bm\beta,t) = \rho_{\bm x}({\bf x},t)\rho_\beta(\bm \beta|\bm x,t)
\end{equation}
The continuity equation is then given by 
\begin{equation}
\frac{\partial  \left(\rho_{\bm x}({\bf x},t)\rho_\beta(\bm \beta|\bm x,t)\right)}{\partial t} = -\nabla \cdot \bm J(\bm x,s,\bm\beta,t).
\end{equation}
Simple integration with respect to $\bm\beta$ yields the reduced continuity equation 
\begin{equation}
\frac{\partial \rho_{\bm x}(\bm x,t)}{\partial t}= -\nabla \cdot \bm J(\bf x,s,\langle \bm\beta|\bm x \rangle,t).\label{eqnpde0}
\end{equation}
This step is valid for all the non-dimensionalized models we consider as they are all linear in their dimensionless parameters (see \cite{Touboul2008}).   The flux has also been redefined upon integration to 
$$(J^V,J^W) = \rho_{\bm x}(\bm x,t) \left(G_1(\bm x,\langle\bm\beta|x\rangle,s),G_2(\bm x,\langle\bm\beta|x\rangle) \right).$$
We now apply the moment closure assumption $\langle \bm\beta|\bm x \rangle = \langle \bm\beta \rangle$ to yield the following PDE: 
\begin{equation}
\frac{\partial \rho_{\bm x}(\bm x,t)}{\partial t}= -\nabla \cdot \bm J(\bf x,s,\langle \bm\beta\bm  \rangle,t).\label{eqnpde}
\end{equation} 
It should be clear that this is equivalent to the continuity equation for a 
homogeneous network with parameter values fixed at $\langle \bm \beta \rangle$. 
Thus, the associated mean-field system is identical to the homogeneous case, only with the parameters fixed at $\langle\bm \beta\rangle$.  This is the simplest assumption one can make in the heterogeneous case.    For example, if we treat $I$ as the source of heterogeneity for a network of Izhikevich neurons, with distribution $\rho_I(I)$, then the resulting mean-field system is
\begin{eqnarray}
\langle w\rangle'&=& \frac{b\langle v\rangle - \langle w\rangle}{\tau_w} + w_{jump}\langle R_i(t)\rangle \label{wone2} \\
s'&=& -\frac{s}{\tau_s} + s_{jump}\langle R_i(t)\rangle  \label{stwo2} \\
\langle R_i(t) \rangle &=&    \left\{ \begin{array}{cl}  
\left( \int_{V}\frac{dv}{v(v-\alpha)-\langle w \rangle + \langle I \rangle+ g(e_r -v)s}\right)^{-1} & : H(\langle w \rangle,s,\langle I \rangle)\geq 0 \\      
0 &: H(\langle w \rangle,s,\langle I \rangle)< 0    
\end{array}   \right.\label{switch12}  \\
H(\langle w \rangle,s,\langle I \rangle) &=& \langle I\rangle - \langle w \rangle -\frac{(\alpha+gs)^2}{4} + ge_rs \label{switch22}\\
\langle v \rangle &=&  \left\{  \begin{array}{lr}  \frac{\langle R_i(t)\rangle}{2}\log\left(\frac{(v_{peak} - \frac{\alpha+gs}{2})^2 +H(\langle w \rangle,s,\langle I \rangle)}{(v_{reset} - \frac{\alpha+gs}{2})^2 +H(\langle w \rangle,s,\langle I \rangle)}\right) + \frac{\alpha+gs}{2}   & : H(\langle w \rangle,s,\langle I \rangle)\geq 0  \\      \frac{\alpha+gs}{2} - \sqrt{-H(\langle w \rangle,s,\langle I \rangle)}  &:H(\langle w \rangle,s,\langle I \rangle)<0   \end{array}   \right. \label{vm1}
\end{eqnarray}
Note that $I$ in equations \eqref{Izswitch1}-\eqref{HIz} has been replaced by $\langle I \rangle$ in equations (\ref{switch12})-(\ref{vm1}).    We treat this system as the baseline mean-field model for comparison purposes, in addition to direct numerical simulations of the network, and we denote this system of equations as mean-field one (MFI).  We should expect this system to be an adequate approximation to the actual network for narrowly centered distributions of the parameter heterogeneity (small values of the variance, $\sigma_\beta$).  

This set of differential equations is representative of a common 
approach taken when fitting actual neurons.  In this approach, multiple 
estimates of parameters or measurements taken from multiple neurons are 
averaged to yield a single parameter value, which is really the mean 
parameter value, $\langle \bm\beta\rangle$.  Simulations of homogeneous, 
large networks are then run with the parameters fixed at their mean values. 
As we shall see in subsequent sections, the behavior of a simulated heterogeneous
network can differ substantially from that of MFI.

\subsubsection{Mean-Field System II}
To derived our second mean-field system, we begin by writing the density 
function in the alternative conditional form 
\begin{equation}
\rho(v,w,\bm\beta,t) = \rho_{W}(w,t|\bm\beta,v)\rho_V(v,t|\bm\beta)\rho_\beta(\bm\beta).
\end{equation}
Next we integrate the continuity equation with respect to $w$.  This yields the following system 
\begin{eqnarray}
\frac{\partial \rho_V(v,t|\bm\beta)}{\partial t}\rho_\beta(\bm\beta) &=& -\int_W \left(\frac{\partial J^V(v,w,s,\bm\beta,t)}{\partial v} + \frac{\partial J^W(v,w,s,\bm\beta,t)}{\partial w}\right) \,dw  \nonumber\\
&=& -\frac{\partial}{\partial v}J(v,\langle w | v,\bm\beta \rangle,s,\bm\beta,t) -  {J^W(v,w,s,\bm\beta,t)|_{\partial W} }\nonumber\\
&=& -\frac{\partial}{\partial v}J(v,\langle w | v,\bm\beta \rangle,s,\bm\beta,t),\label{MFIIPDE} 
\end{eqnarray}
where the last term vanishes as $J^W$ is assumed to be vanishing on the 
boundary, and 
\begin{equation}
J(v,\langle w | v,\bm\beta \rangle,s,\bm\beta,t)  = \int_W J^V(v,w,s,\bm\beta,t)\,dw .
\end{equation}

We now make the first order moment closure assumption $\langle w | v,\bm\beta\rangle = \langle w \rangle$.  Then to complete the system, we must derive a differential equation for $\langle w \rangle$: 
\begin{eqnarray}
\langle w \rangle'& =& \int_V\int_W\int_{\bm\beta}w \frac{\partial \rho(v,w,\bm\beta,t)}{\partial t}\,d\bm\beta\, dw\,dv \nonumber \\
&=&-\int_V\int_W \int_{\bm\beta}  w\left( \frac{\partial J^W}{\partial w} + \frac{\partial J^V}{\partial v}\right)\,d\bm\beta\, dw\,dv \nonumber \\
&=& \int_V\int_W\int_{\bm\beta} G_2(v,w,\bm\beta) \rho(v,w,\bm\beta,t) \,d\bm\beta\, dw\,dv 
 \nonumber\\& -&  \int_W\int_{\bm\beta} w(J^V(v_{peak},w,s,\bm\beta,t) - J^V(v_{reset},w,s,\bm\beta,t)) \,d\bm\beta\, dw \nonumber\\
&=& \langle G_2(v,w,\bm\beta)\rangle  -   \int_W\int_{\bm\beta} w(J^V(v_{peak},w,s,\bm\beta,t) - J^V(v_{peak},w-w_{jump},s,\bm\beta,t) )\,d\bm\beta\, dw \label{BC2}\\
&=& \langle G_2(v,w,\bm\beta)\rangle +\int_{\bm\beta}\int_W w_{jump} J^V(v_{peak},w,s,\bm\beta,t)\,dw\, d\bm\beta  + O(w_{jump}^2) \nonumber\\
&\approx& G_2(\langle v\rangle,\langle w\rangle,\langle\bm\beta\rangle) +\int_{\bm\beta} w_{jump} J(v_{peak},\langle w\rangle,s,\bm\beta,t)\, d\bm\beta. 
\end{eqnarray}
Note that we have made the approximation $\langle G_2(v,w,\bm\beta)\rangle =G_2(\langle v\rangle,\langle w\rangle,\langle\bm\beta\rangle)$ in addition to dropping the $O(w_{jump}^2)$ terms.   Additionally, the substitution on line \eqref{BC2} comes from the boundary condition \eqref{BC1}.  

Applying a quasi-steady state approximation to
the PDE \eqref{MFIIPDE} yields the following equation for the steady state voltage independent flux, $J(s,\langle w \rangle,\bm\beta)$:
\begin{equation}
J(s,\langle w \rangle,\bm\beta)=\begin{cases} \left[\int_{V}\frac{dv}{G_1(v,s,\langle w\rangle,\bm\beta)} \right]^{-1}\rho_\beta(\bm\beta) &\text{if} \quad H(\langle w \rangle,s,\bm\beta)\geq 0\\ 0 &\text{if}\quad  H(\langle w \rangle,s,\bm\beta)<0 \end{cases}.
\label{Jbeta}\end{equation}
We interpret the ratio $J(s,\langle w\rangle,\bm\beta)/\rho_{\bm\beta}(\bm\beta)$ as the parameter dependent (or conditional) network averaged firing rate, $\langle R_i(t)|\bm\beta\rangle$,
based on the fact that
$$\int_{\bm\beta} J(s,\langle w\rangle ,\bm\beta)\, d\bm\beta \approx \langle R_i(t)\rangle.$$ 
In other words, the distribution of parameters induces a distribution of firing rates across the network, and the network averaged firing rate is the mean of the distribution.  

In summary, the resulting mean-field equations are given by: 
\begin{eqnarray}
\langle w\rangle'&=& \frac{b\langle v\rangle- \langle w\rangle}{\tau_w} +\int_{\bm\beta} w_{jump}\langle R_i(t)|\bm\beta\rangle\rho_\beta(\bm\beta)\,d\bm\beta \label{wwone} \\
s'&=& -\frac{s}{\tau_s} + s_{jump}\int_{\bm\beta}\langle R_i(t)|\bm\beta\rangle\rho_\beta(\bm\beta)\,d\bm\beta \label{sstwo} \\
\langle R_i(t)|\bm\beta \rangle &=&    \left\{     \begin{array}{lr}  \left[\int_{V}\frac{dv}{G_1(v,s,\langle w\rangle,\bm\beta)} \right]^{-1}& : H(\langle w \rangle,s,\bm\beta)\geq 0  \\      0 &: H(\langle w \rangle,s,\bm\beta)<0      \end{array}   \right.\label{sswitch1}  \\
 H(\langle w \rangle,s,\bm\beta) &=&I -\langle w \rangle  +  \min_{v}(F(v) + g(e_r - v)s) \label{HMFII}\\
\langle v\rangle &=& \int_{\bm\beta}\langle v|\bm\beta \rangle \rho_{\beta}(\bm\beta)\,d\bm\beta\label{vm2}
\end{eqnarray}
where the forms of $G_1(v,s,\langle w \rangle,\bm\beta)$ and $H(\langle w \rangle,s,\bm\beta) $ depend on which specific neural model is used, and the equation for $\langle v|\bm\beta\rangle$ can be found in Appendix A. 
Note that the distribution of firing rates is not computed explicitly in these equations, only the conditional firing rates, $\langle R_i(t) |\bm\beta\rangle$, are 
computed. 
However, we show in section~\ref{mf3sec} that a distribution for the steady state firing rates of the network can be computed using $\langle R_i(t)|\bm\beta\rangle$.  

We refer to equations \eqref{wwone}-\eqref{vm2} as mean-field two (MFII).  It should be noted that MFI and MFII effectively differ in the order in which the integrations are carried out.  In MFI, we integrate with respect to $\bm\beta$ first, and then apply the first order moment closure assumptions $\langle \bm\beta|\bm x\rangle = \langle \bm\beta\rangle$ and $\langle w | v \rangle = \langle w \rangle$.  In MFII, we integrate with respect to $w$ first, and then apply the moment closure assumption $\langle w | v,\bm\beta\rangle = \langle w\rangle$.   Furthermore, if $\langle R_i(t)| \bm\beta\rangle$ does not actually depend on the heterogeneous parameter $\bm\beta$, such as when the heterogeneity is in $w_{jump}$, then MFI and MFII are identical.   

In fact the first order moment closure assumption used here can be 
weakened.  This leads to the ``mean-field" system  in the next subsection, which is a different kind of system than MFI and MFII.     

\subsubsection{Mean-Field System III} 
Suppose that instead of assuming that $\langle w | v,\bm\beta\rangle = \langle w \rangle$, we make the weaker assumption that $\langle w | v,\bm\beta\rangle = \langle w | \bm\beta \rangle$.  It turns out that this assumption yields a PDE, even when one makes the quasi-steady state approximation, as we now show.    
Applying this weaker moment closure assumption to \eqref{MFIIPDE} yields 
the following simplification of the continuity equation: 
\begin{equation}
\frac{\partial \rho_V(v,t|\bm\beta)}{\partial t}\rho_\beta(\bm\beta)= -\frac{\partial}{\partial v}J(v,\langle w | \bm\beta \rangle,s,\bm\beta,t). 
\end{equation}
Application of the quasi-steady state approximation now yields 
\begin{eqnarray*}
J(v,s,\langle w|\bm\beta \rangle,\bm\beta) &=&    \left\{     \begin{array}{lr}  \left[\int_{V}\frac{dv}{G_1(v,s,\langle w|\bm\beta \rangle,\bm\beta)} \right]^{-1}\rho_\beta(\bm\beta)& : H(\langle w|\bm\beta \rangle,s,\bm\beta)\geq 0  \\      0 &: H(\langle w|\bm\beta \rangle,s,\bm\beta)<0      \end{array},   \right.\label{sswitch12}  \\
 H(\langle w|\bm\beta \rangle,s,\bm\beta) &=&I -\langle w|\bm\beta \rangle  +  \min_{v}(F(v) + g(e_r - v)s).
\end{eqnarray*}

An equation for the time variation of $\langle w | \bm\beta \rangle$   
can be derived in a similar manner to the last section, yielding
the following mean-field system: 
\begin{eqnarray}
\langle w | \bm\beta\rangle' &=& \frac{b\langle v|\bm\beta\rangle-\langle w | \bm\beta\rangle}{\tau_w} + w_{jump}\langle R_i(t)|\bm\beta\rangle \label{wone3} \\
s'&=& -\frac{s}{\tau_s} + s_{jump}\int_\beta\langle R_i(t)|\bm\beta\rangle\rho_\beta(\bm\beta)\,d\bm\beta \label{stwo3} \\
\langle R_i(t)|\bm\beta \rangle &=&    \left\{     \begin{array}{lr}  \left[\int_{V}\frac{dv}{G_1(v,s,\langle w|\bm\beta\rangle,\bm\beta)} \right]^{-1}& : H(\langle w|\bm\beta\rangle,s,\bm\beta)\geq0     \\      0 &:H(\langle w|\bm\beta\rangle,s,\bm\beta)<0     \end{array}   \right.\label{MF3R} 
\end{eqnarray}
Note 
that $\langle w \rangle$ can be computed via:
\begin{equation}
\langle w \rangle = \int_\beta \langle w | \bm\beta\rangle \rho_\beta(\bm\beta)\,d\bm\beta. 
\end{equation}
We denote this system as mean-field three (MFIII).  
Note that the equation for $\langle w|\bm\beta \rangle$ is actually a PDE. 
Thus, while this system should be more accurate than mean-field II, it has 
the drawback of being more difficult to analyze.  The dependence on 
$\bm\beta$ forces one to discretize over a mesh in $\bm\beta$ in order to 
work numerically with this system.  This makes numerical bifurcation analysis more 
difficult.   However, as we shall show later, an approach that yields similar information to direct bifurcation analysis can be used with MFIII.  
The equation for $\langle v|\bm\beta\rangle$ can once again be found in the appendix.  

\section{Results} 
\subsection{Numerical Simulations}
Recall that simulations of a homogeneous network and the corresponding
mean-field system are shown in Figure~\ref{fig1}. Note that the network
undergoes a bifurcation from tonically firing to bursting as the amount of 
applied current $I_{app}$ is decreased, with all other parameter values held fixed.
Further simulations show that if $I_{app}$ is decreased below $I_{rh}$ then all 
neurons in the network are quiescent (non-firing). 

To determine and compare the validity of the three mean-field systems, we 
have run a series of numerical simulations of these systems and of an actual 
network containing 1000 neurons.  The parameter values for the individual 
neurons can be found in Table~\ref{table1}.  They are based on those given 
in~ \cite{us} which were fit to data for hippocampal CA3 pyramidal neurons 
from~ \cite{Hemond}. 
These are the parameter values we use for the rest of this paper, unless 
otherwise indicated.  

To begin, we consider heterogeneity only in the applied current. The 
distributions are assumed to be normal with mean $\langle I \rangle$ and 
standard deviation $\sigma_I$. We varied the values of the mean and standard
deviation and found that the accuracy of the mean-field approximations 
depends on where the mean is relative to the different bifurcation 
regions and on the size of the standard deviation. 

\begin{table*}[htp]
\centering
\begin{tabular}{| l || l | l || l | }
  \hline
   \multicolumn{2}{|c| }{Dimensional Parameters}&    \multicolumn{2}{|c| }{Dimensionless Parameters} \\
   \hline                        
   $C$& 250 pF  & & \\
    $k$ & 2.5  nS/mV & &\\
   $V_R$ & -65  mV & &\\
   $V_T$ & $V_R + 40 - \frac{b}{k} = 41.7$mV & $\alpha = 1+ \frac{V_T}{|V_R|}$ & 0.6215  \\
   $V_{peak}$ & 30 mV & $v_{peak}=1+\frac{V_{peak}}{|V_R|}$ & 1.461\\ 
   $V_{reset}$& -55 mV& $v_{reset}=1+\frac{V_{reset}}{|V_R|}$ &0.1538\\ 
   $W_{jump}$ &200 pA& $w_{jump}=\frac{W_{jump} }{k |V_R|^2}$ & 0.0189\\
   $\tau_W$  & 200 $\text{ms}$ & $a= \left(\frac{\tau_W k |V_R|}{C}\right)^{-1}$ & 0.0077\\
   $\eta$ & -1 nS &$b = \frac{\eta}{k|V_R|}$ &  -0.0062\\  
   $I_{app}$  & 1000 - 5000 pA &$I = \frac{I_{app}}{k|V_R|^2}$& 0.0776 - 0.3333 \\ 
  $g_{syn}$ & 0 - 600 nS & $g = \frac{g_{syn}}{k|V_R|}$ & 0 - 3.6923 \\
  $\tau_{syn}$ & 4 ms& $\tau_s=\frac{\tau_{syn} k |V_R|}{C}$ & 2.6\\
  $s_{jump}$ & 0.8 & & \\
  $N$ & 1000  & &\\ 
  $\sigma_I$ & 0 - 1000 pA & &\\ 
  $\sigma_g$ & 50 nS  & &\\ 
  $\sigma_d$ & 50 pA  & &\\ 
   $m$ (mixing parameter) & 0 - 1  & &\\ 
      \hline  
 \end{tabular}
\caption{The parameters and distribution variances used in this paper.   These parameters apply unless otherwise indicated. Rheobase for the dimensional parameter values is $I_{rh}=1000$ pA.} 
\label{table1}
\end{table*}

\begin{figure}
\centering
          \subfigure[$\meanIapp=5000$ pA,  $\sigma_I = 2000$ pA]{\includegraphics[scale=0.5]{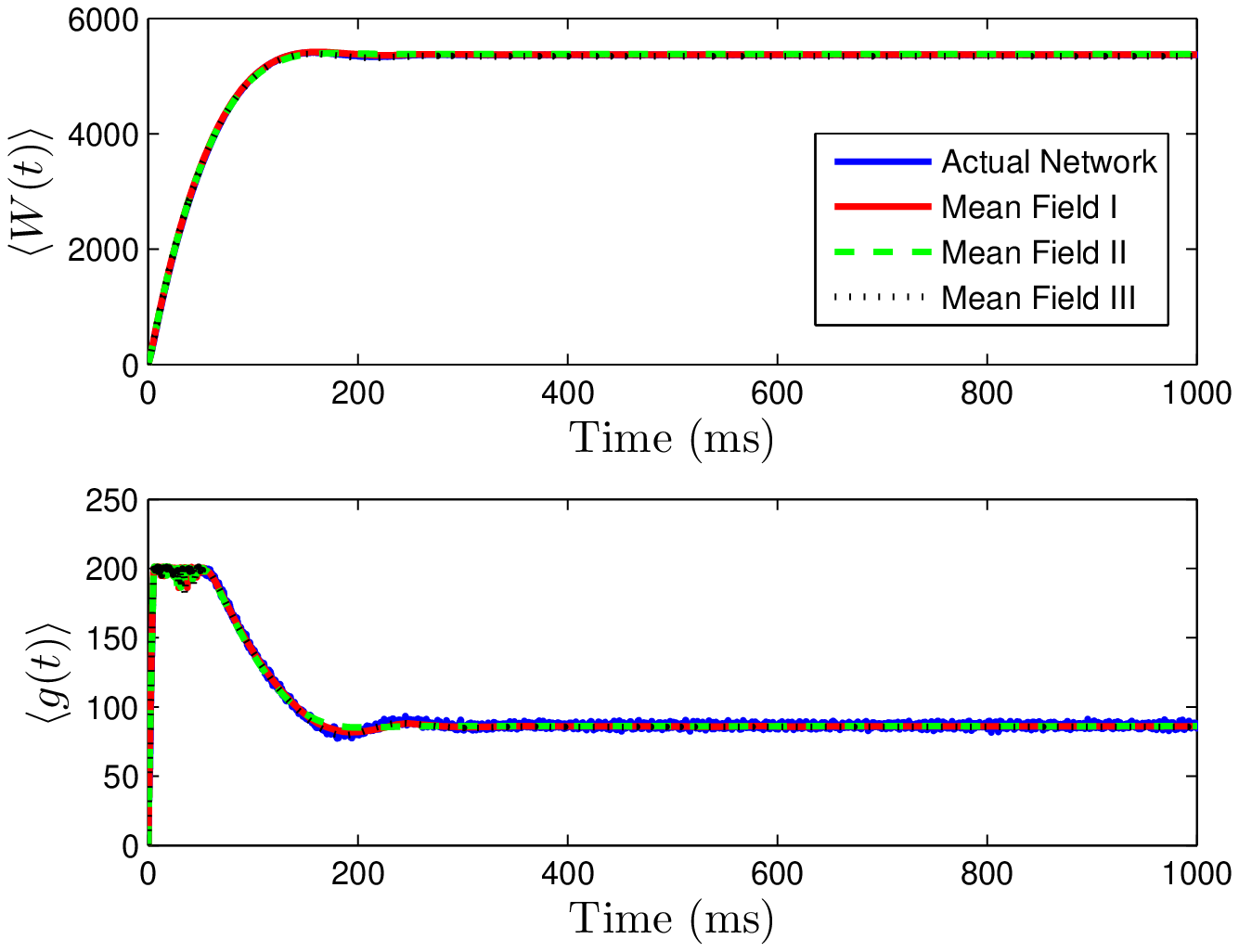}
\label{fig2a}}
         \qquad 
           \subfigure[$\meanIapp=3000$ pA,  $\sigma_I = 200$ pA]{\includegraphics[scale=0.5]{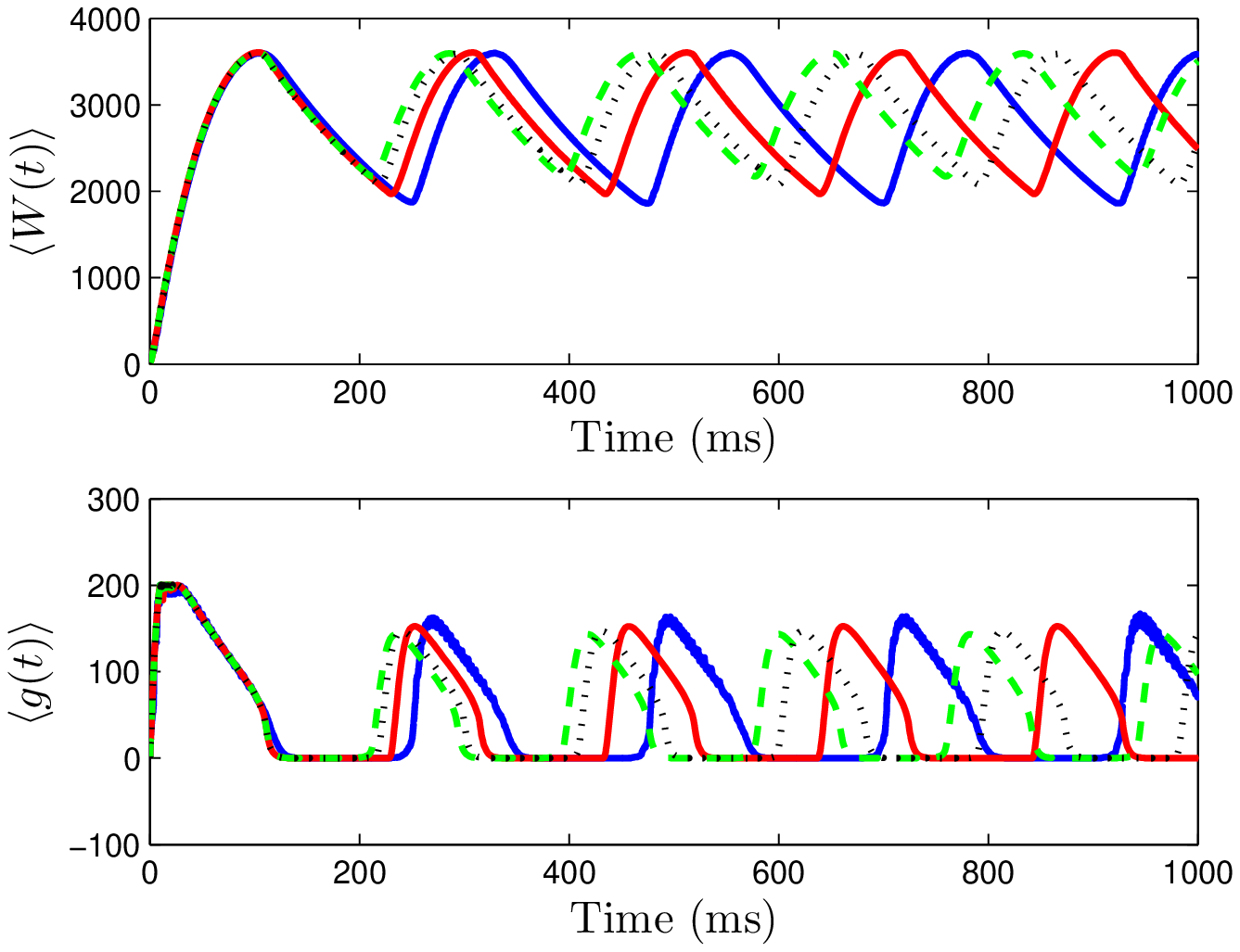}
\label{fig2b}}
          \\ 
          \subfigure[$\meanIapp=3000$ pA,  $\sigma_I = 500$ pA]{\includegraphics[scale=0.5]{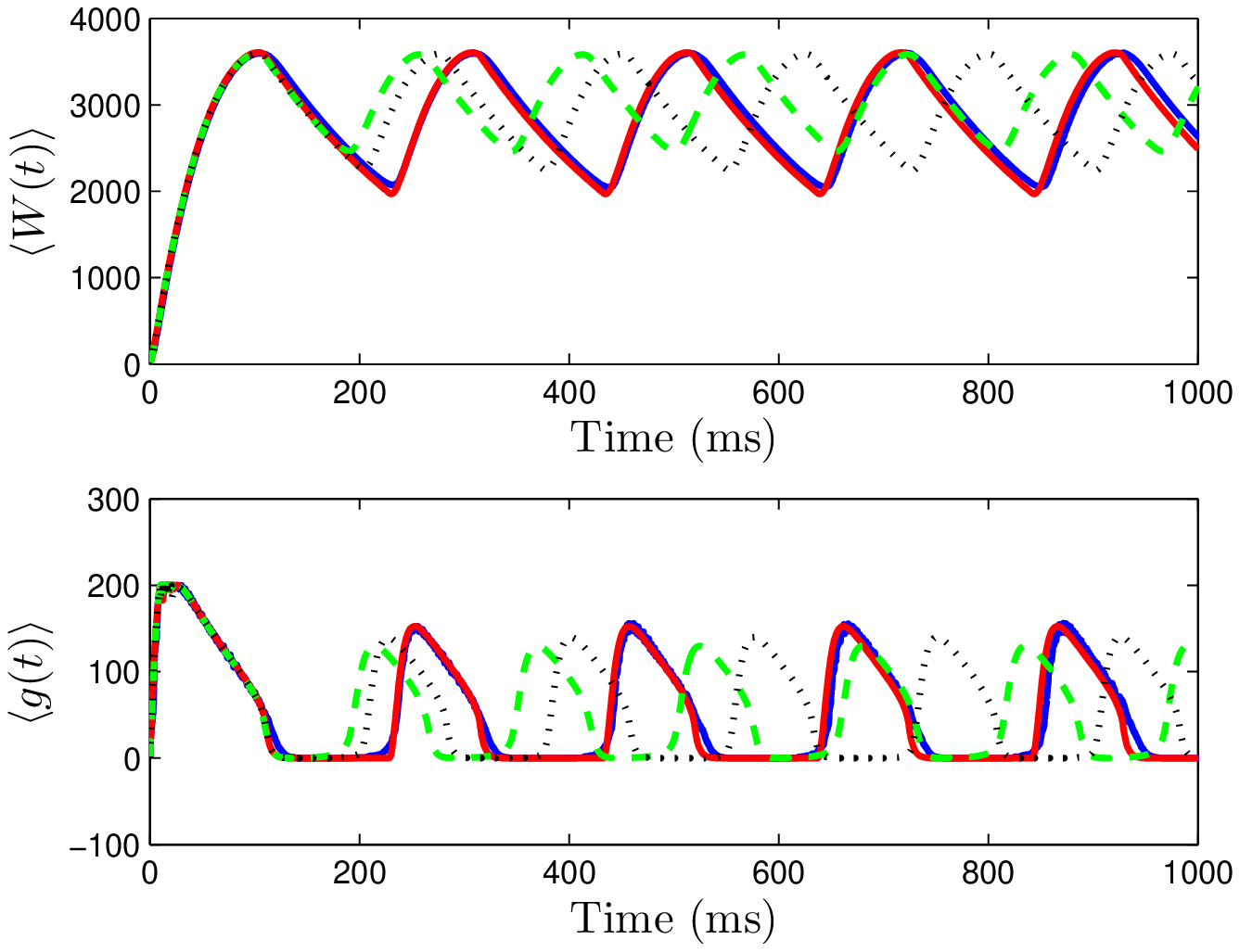}
\label{fig2c}}
         \qquad 
           \subfigure[$\meanIapp=3000$ pA,  $\sigma_I = 2000$ pA]{\includegraphics[scale=0.5]{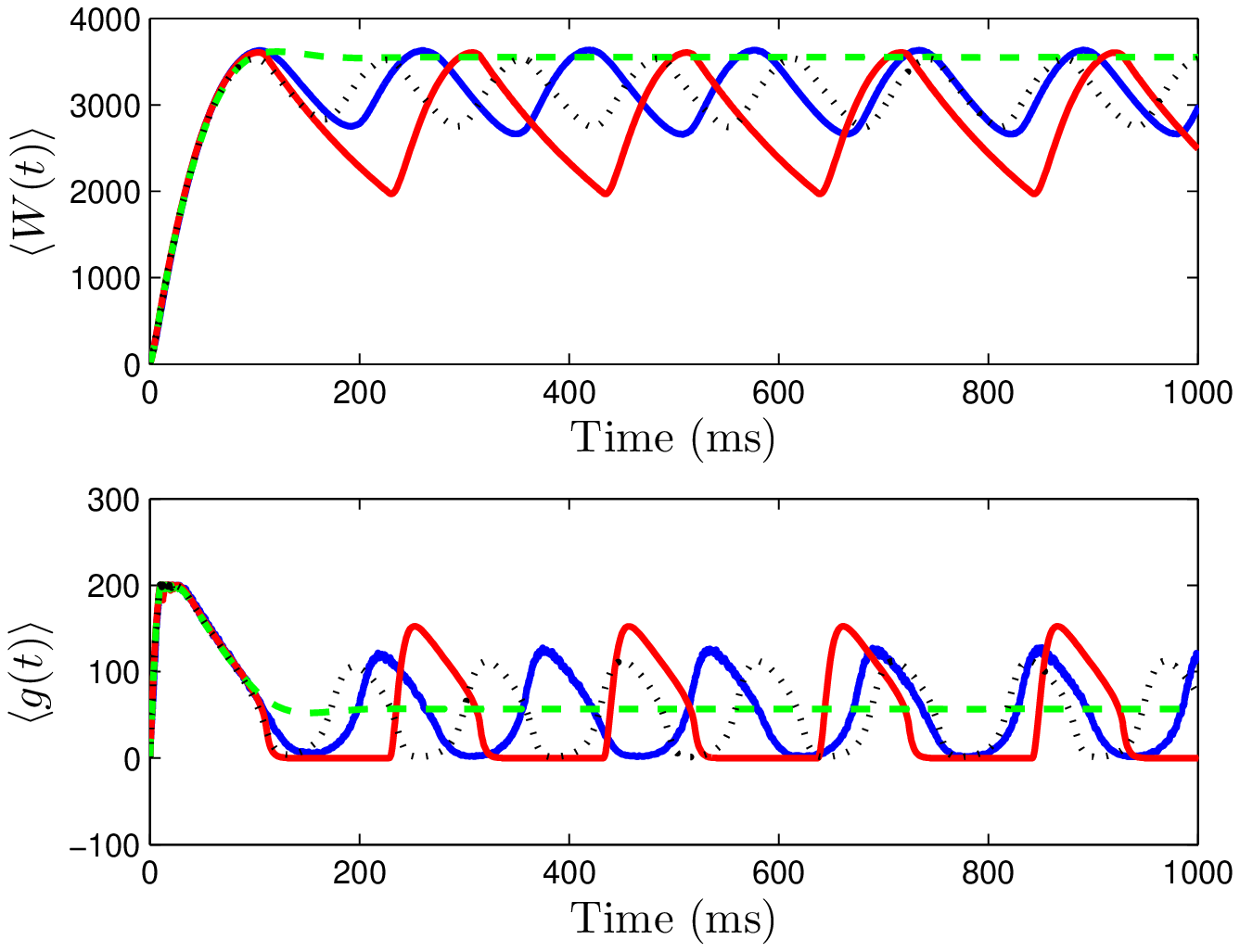}
\label{fig2d}}
        \caption{
Numerical simulations of a network of 1000 Izhikevich neurons with parameters as in Table 1, except $g_{syn}=200$ and the applied current which is normally distributed with mean and variance as shown. Blue is the network average of a given variable, red is MFI, green is MFII and black is MFIII.  
In this region, the mean-driving current is away from rheobase, $\meanIapp \gg I_{rh}$. All three approximations are quantitatively and qualitatively similar for small to intermediate sized variances in the distribution of currents.  For small variances, MFI is the most accurate and for larger variances, MFIII is the most accurate.  For large variance, MFII bifurcates back to tonic firing earlier then MFI and MFIII, as seen in (d)} \label{fig2}
\end{figure}

As for the homogeneous network, the heterogeneous network undergoes a bifurcation 
from tonic firing to bursting as the amount of current applied to the individual neurons
is decreased, with all other parameters held fixed. This can be seen in Figure 
\ref{fig2} where the bifurcation with decreasing $\meanIapp$ is shown.  As $\meanIapp$ 
is decreased below $I_{rh}$, there is a 
bifurcation to quiescence. 
We will not discuss this latter bifurcation in detail, as we are primarily interested in analyzing the transition from tonic firing to 
bursting.  

Note that the bifurcations described above only occur in the mean sense. Since 
the current values are normally distributed, there is non-zero probability that 
some neurons receive large enough or small enough current to be in a state other 
than that corresponding to the value of $\meanIapp$.  For small enough standard 
deviations, very few neurons in an actual finite network are likely to have
behaviour different from the mean.  However, for large standard deviations, 
a sizable proportion may not follow the mean behaviour.

Given this knowledge of the different qualitative behaviors of the network, we can see how the mean-field systems compare.  For tonic firing (Figure \ref{fig2a}), even when the standard deviation is large, the mean-field systems approximate the network means $\langle g(t)\rangle$, and $\langle W(t)\rangle$ very well.  However,  when the network is bursting, with $\langle I_{app} \rangle >I_{rh}$, we see a difference as to which mean-field system is superior.  For small values of $\sigma_I$, we have numerically found that mean-field I is superior to mean-field II and III, however all the systems are quantitatively and qualitatively accurate (see Figure \ref{fig2b},\ref{fig2c}).  However, for larger values of $\sigma_I$, the amplitude error of MFIII is the smallest, and MFII is the worst approximation as it bifurcates to tonic firing prematurely.  

\begin{figure}
\centering
          \subfigure[$\meanIapp=1200$ pA,  $\sigma_I = 200$ pA]{\includegraphics[scale=0.5]{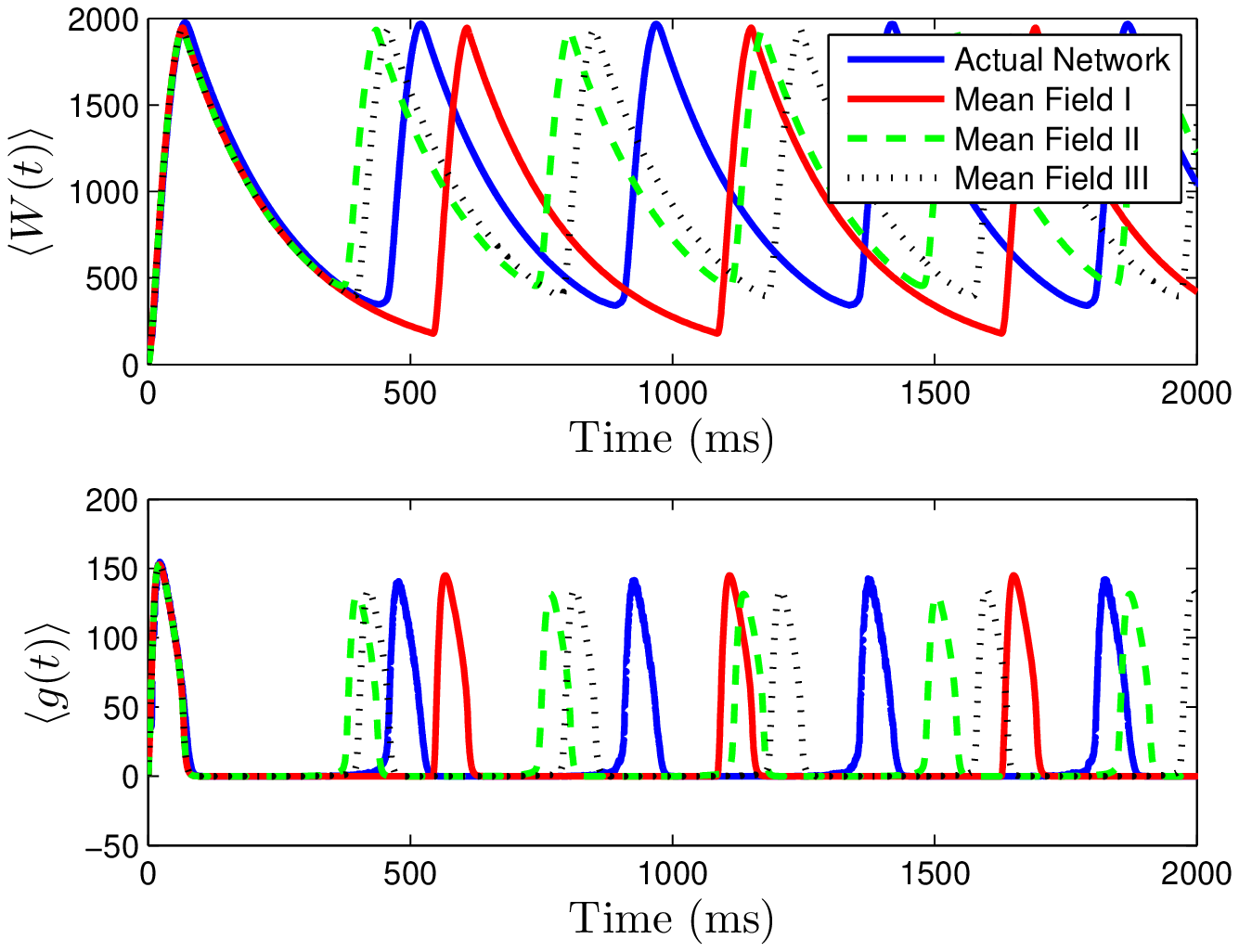}
\label{fig3a}}
         \qquad 
           \subfigure[$\meanIapp=1200$ pA,  $\sigma_I = 500$ pA]{\includegraphics[scale=0.5]{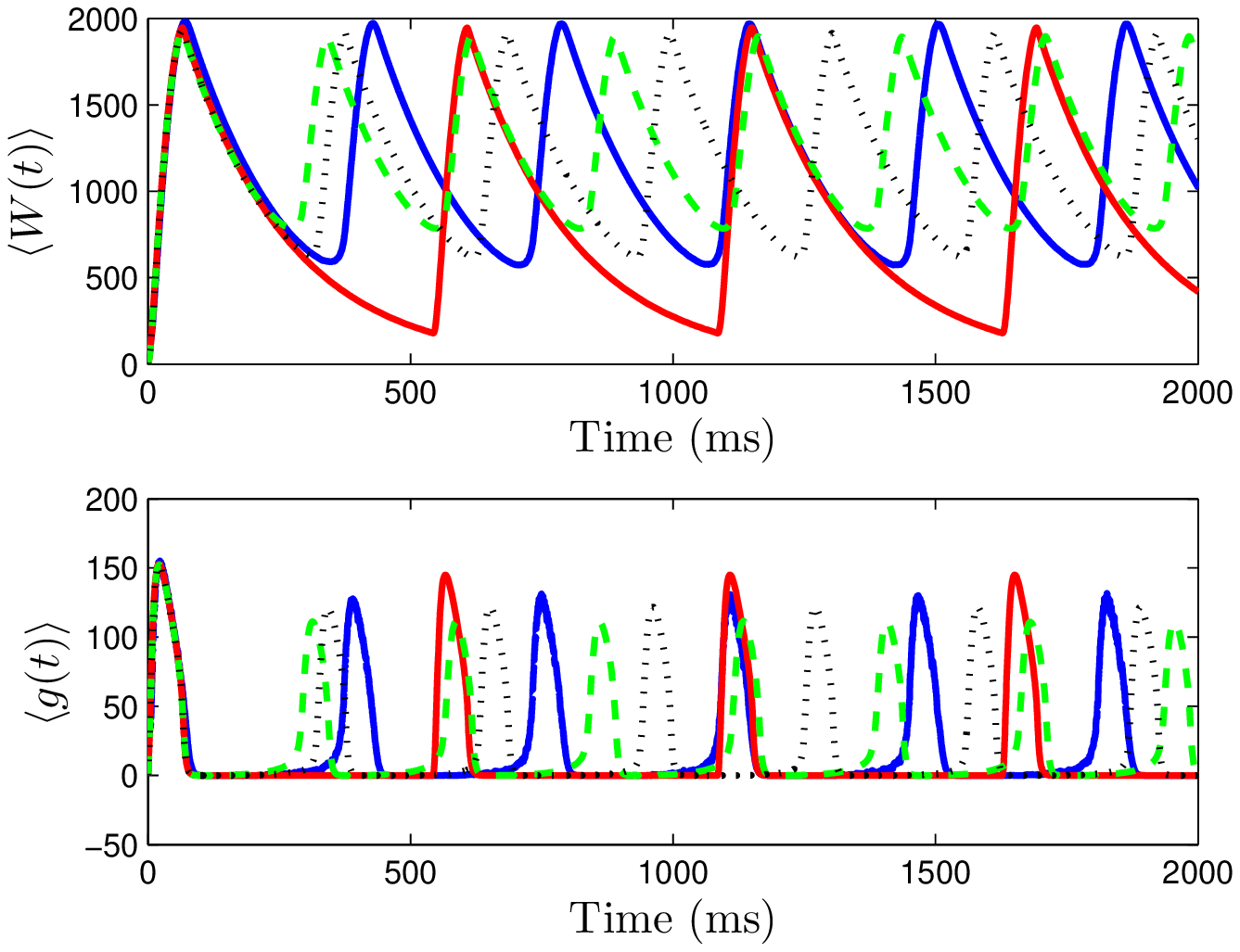}
\label{fig3b}}
          \\ 
          \subfigure[$\meanIapp=1200$ pA,  $\sigma_I = 1000$ pA]{\includegraphics[scale=0.5]{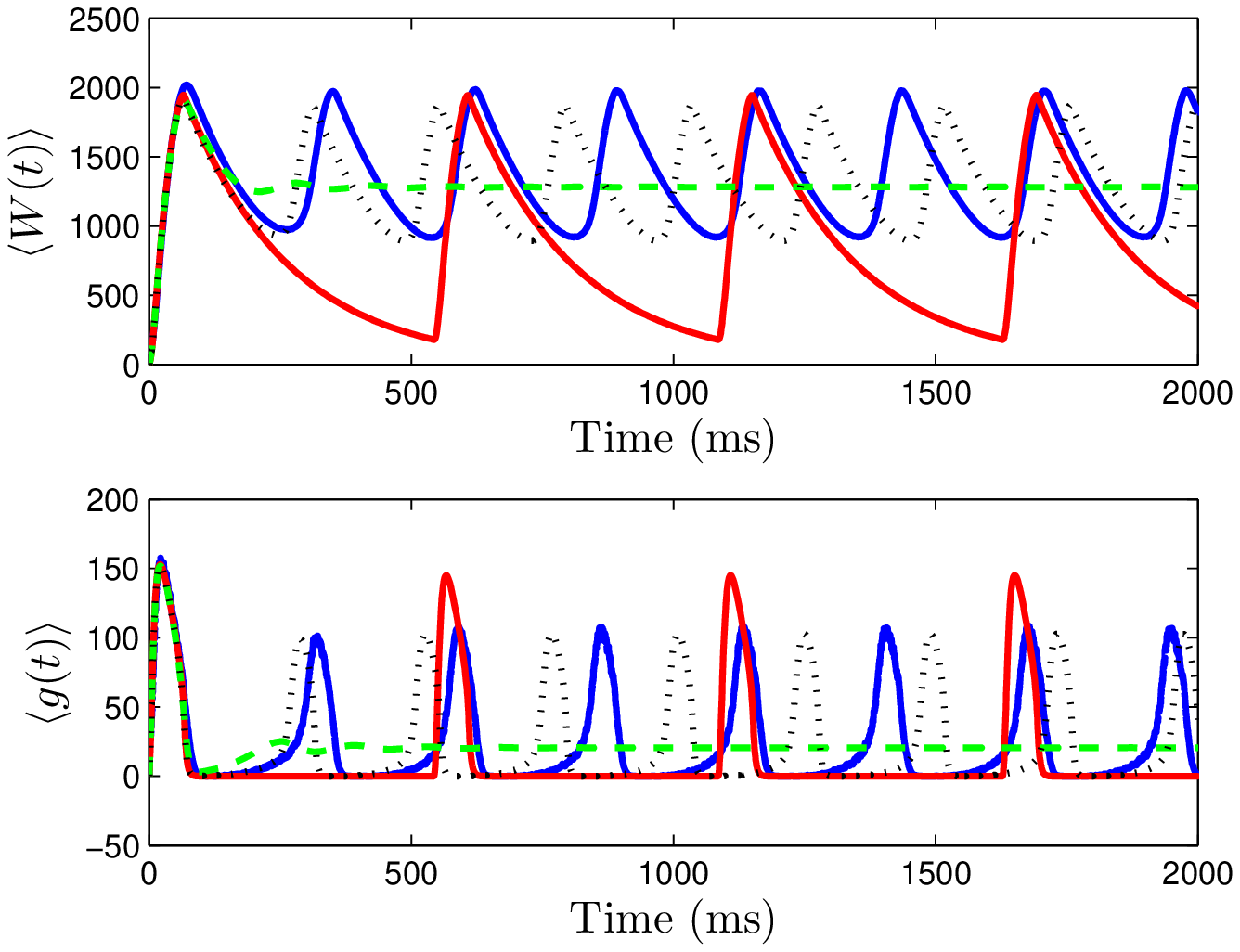}
\label{fig3c}}
         \qquad 
           \subfigure[$\meanIapp=1200$ pA, $\sigma_I = 2000$ pA]{\includegraphics[scale=0.5]{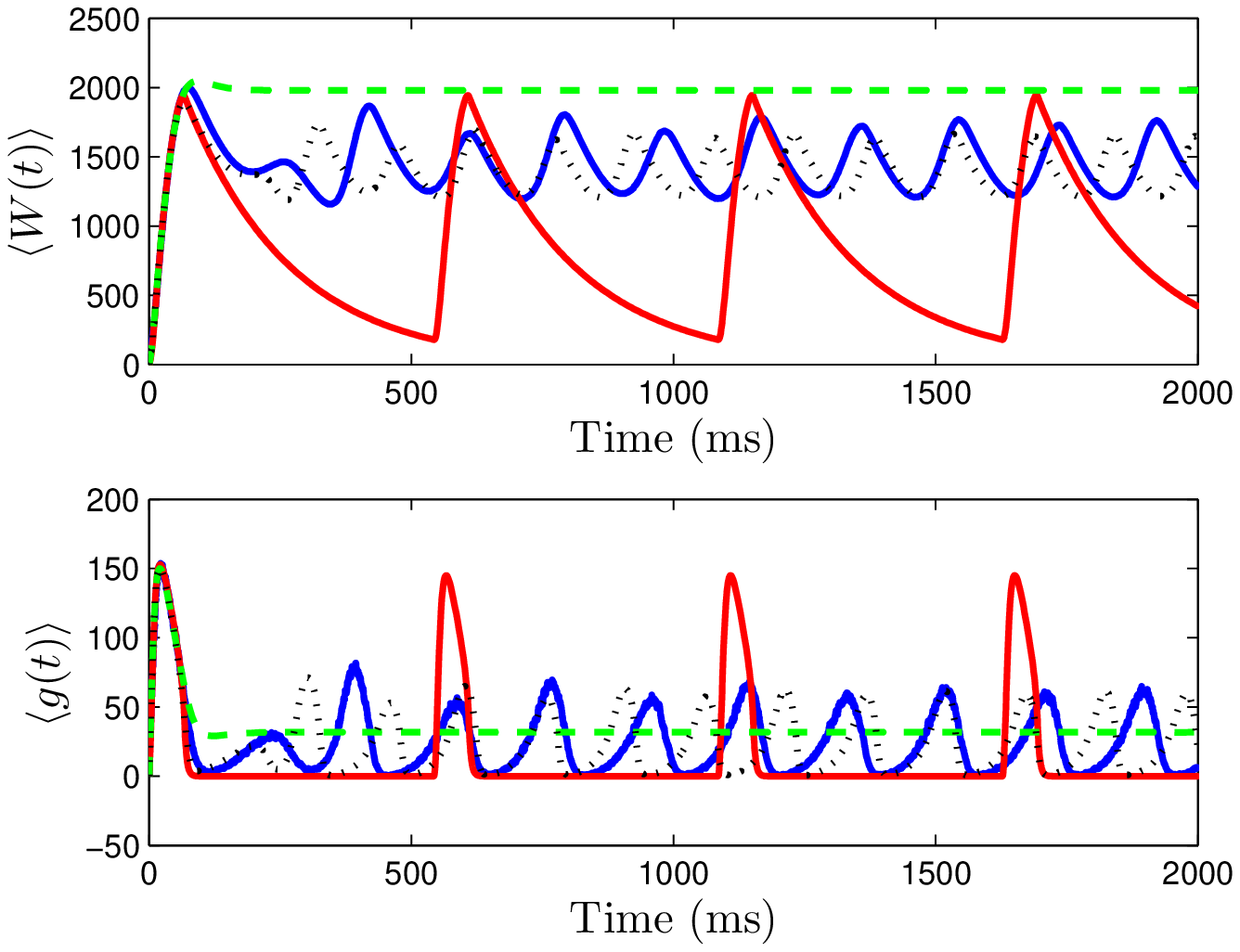}
\label{fig3d}}
        \caption{
Numerical simulations of a network of 1000 neurons with parameters as in Table 1, except $g_{syn}=200$ and the applied current which is normally distributed with mean and variance as shown. Blue is the network average of a given variable, red is MFI, green is MFII and black is MFIII.  
In these simulations, the mean-driving current is close to (and over) the rheobase.  
In all cases, MFI is the least accurate.  This is because it depends 
only on $\meanIapp$. When $\meanIapp = O(I_{rh})$, even for small variance, many of 
the neurons have $I< I_{rh}$ and may not spike at all.  (a),(b) For small 
values of $\sigma_I$, all three approximations are qualitatively and quantitatively accurate.  (c),(d) For larger variance, $\sigma_I = O(I_{rh})$, only MFIII is qualitatively and quantitatively accurate. In this case, MFII bifurcates early to tonic firing. 
} \label{fig3}
\end{figure}

When $\langle I_{app} \rangle$ is close to $I_{rh}$, we see even stronger differences between the three mean-field systems.   For small to intermediate standard deviations, MFII and MFIII are clearly superior to MFI, having a smaller amplitude and frequency error (see Figure \ref{fig3a}, \ref{fig3b}).   However, for larger values of $\sigma_I$ as shown in Figure \ref{fig3c}, and \ref{fig3d}, only MFIII is a qualitatively and quantitatively accurate representation of the behavior of the network.   The amplitude and frequency error of MFI are very large, and MFII again bifurcates prematurely to tonic firing. 

\begin{figure}
\centering
          \subfigure[$\meanIapp=1100$ pA,  $\sigma_I = 2000$ pA]{\includegraphics[scale=0.5]{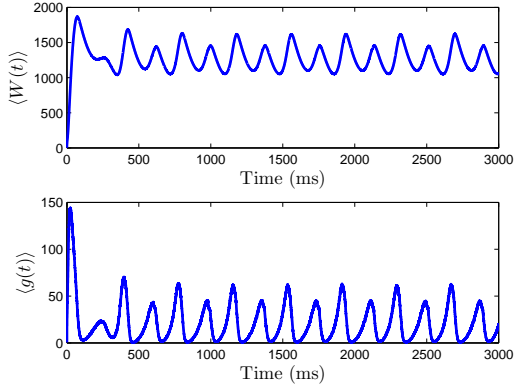}
\label{fig4a}}
         \qquad 
           \subfigure[Raster plot of the simulation]{\includegraphics[scale=0.5]{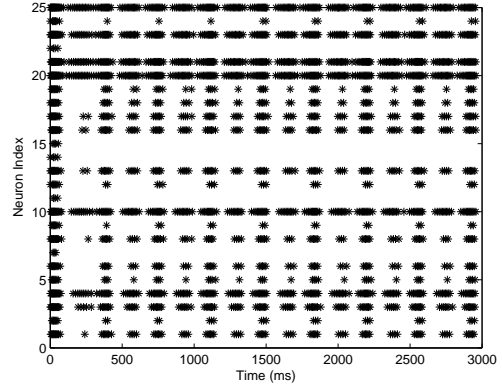}
\label{fig4b}}
          \\ 
          \subfigure[Mean-field equations at same parameter values]{\includegraphics[scale=0.5]{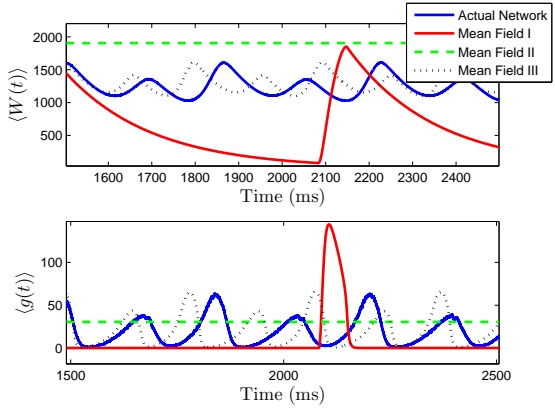}
\label{fig4c}}
         \qquad 
           \subfigure[Steady state period doubled limit cycle]{\includegraphics[scale=0.5]{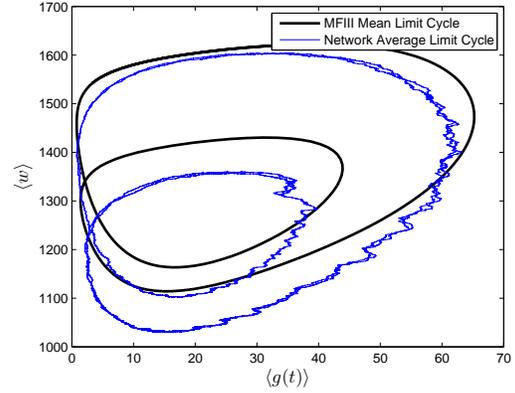}
\label{fig4d}}
        \caption{
Period doubled limit cycle in the heterogeneous network and in MFIII.  The network consists of 5000 neurons, with parameters as in Table 1, except $g_{syn}=200$ and the applied current which is normally distributed with mean and variance as shown in (a).  (a) period-doubled limit cycle for the network shown in terms of the mean variables.  (b) raster plot of 25 randomly selected neurons of the network.  The behaviors include burst firing, alternate burst firing, tonic firing, and quiescence.  (c) numerical simulations of the mean-field systems.  Only MFIII is able to reproduce the period doubling behavior.  (d) Comparison of the ``phase portrait" of period doubled limit cycle for MFIII and the mean variables of network.   
} \label{fig4}
\end{figure}

One should note that for $\langle I_{app} \rangle =O(I_{rh})$ and for large values of $\sigma_I$, the network can undergo a period doubling bifurcation.  This is shown in Figure \ref{fig4}. The large standard deviation in the current forces different neurons into different regimes, such as tonic firing, bursting, alternate burst firing, and quiescence as seen in Figure \ref{fig4b}. The fact that a small subpopulation of neurons are alternate bursters (i.e., burst with twice the 
period of the rest of the bursting neurons) appears as a period
doubled limit cycle in the mean variables of the network, as seen in Figure
\ref{fig4a}. 
Only MFIII is able to approximate the period-doubled limit cycle with any degree of accuracy, as shown in Figure \ref{fig4c} and \ref{fig4d}.  Period doubling bifurcations are well known for their capability of inducing chaos.  Given that MFIII accurately represents the period doubling bifurcation, it may be able to replicate any potential chaotic behavior.  However, we leave further investigation of this interesting behaviour for future work.

To summarize, all the mean-field systems are valid for tonic firing parameter regimes, and MF I is valid for all parameter regimes with small $\sigma_I$, except for $\langle I_{app} \rangle = O(I_{rh})$.  Mean-Field II and III are valid for bursting with $\langle I_{app} \rangle \gg I_{rh}$, and MFIII is the only valid approximation for $\langle I_{app} \rangle = O(I_{rh})$.  Thus, when $\langle I_{app} \rangle$  is large we may be able to use MFII to determine the type of bifurcation(s) involved when a heterogeneous network transitions from tonic firing to bursting and the location in parameter space of the bifurcation manifolds. 
Note that when the mean network behaviour undergoes a bifurcation from a tonic firing steady state to a bursting oscillation, this does not indicate that the entire network of neurons is bursting, or tonically firing.  However, we will show how to use MFIII to determine what proportion of neurons display the different types of behavior, given a specific parameter regime and level of heterogeneity. 

In addition to simple heterogeneity using unimodal distributions, one can also apply the same three mean-field equations to networks where multiple subpopulations exist.  However, unlike previous attempts at modelling networks with multiple subpopulations, we do not generate discrete coupled subnetworks with different fixed values of the parameters in each subnetwork.  Instead we use a smoother  approach where the networks have distributions of parameters with multiples modes indicative of multiple subpopulations.  This can be easily done through the processing of mixing unimodal distributions (see Appendix C).  

\subsection{Applications of Mean-Field Theory with a Single Source of Heterogeneity}
\subsubsection{Bifurcation Types and Manifolds Using MFII}
As shown in Fig.~\ref{fig2} the CA3 model  network a makes 
transition from tonic firing to bursting as $\meanIapp$ is varied. Similar 
transitions occur when $g_{syn}$ is varied.  In this section, we use numerical 
bifurcation analysis of MFII to determine the bifurcations involved in this 
transition, and the manifolds where they occur in the $\meanIapp$-$g_{syn}$ 
parameter space.  Since the mean-field system \eqref{wwone}--\eqref{vm2} 
consists of switching ODEs, this involves bifurcations of non-smooth systems  
as well as standard (smooth) bifurcations.  A review of the theory of non-smooth
systems can be found in \cite{nonsmooth}.  The numerical bifurcation analysis
is done in MATLAB \cite{MATLAB} using the MATCONT package \cite{MATCONT} for
the standard bifurcations and direct numerical simulations for the non-smooth 
bifurcations.  We compare the mean field theory results to those for the
homogeneous system and to direct simulations of large networks. 

\begin{figure}
\centering
\subfigure[MFI, $I_{app} = 2000$ pA]{\includegraphics[scale=0.8]{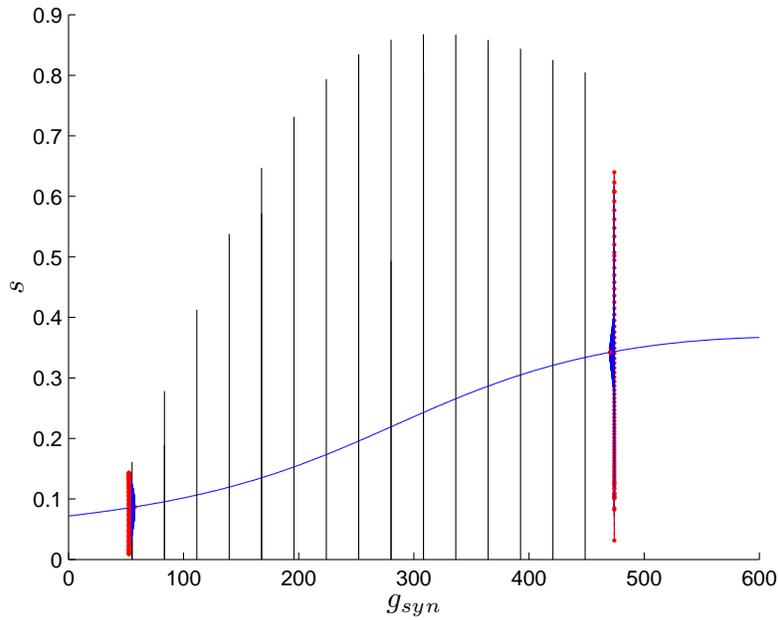}
\label{fig5a}} 
\subfigure[MFII, $\meanIapp= 2000$ pA, $\sigma_I = 500$ pA ]{\includegraphics[scale=0.8]{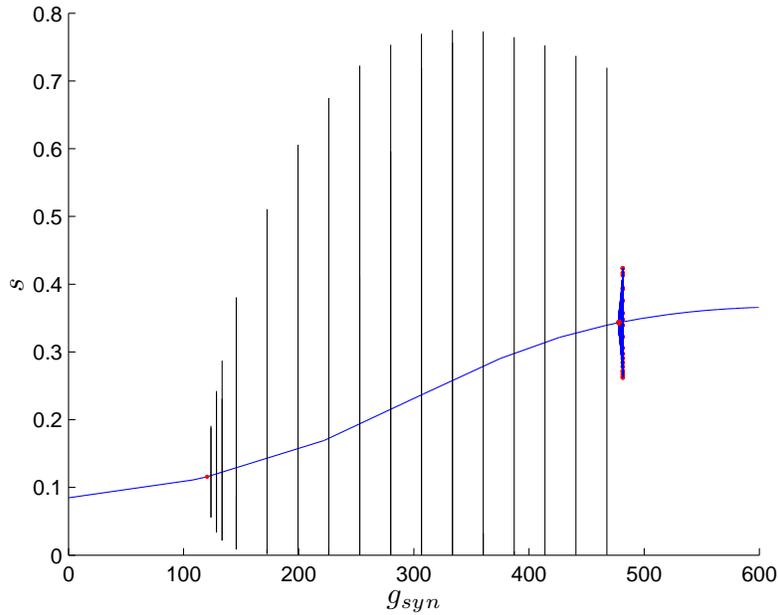}
\label{fig5b}}
   \caption{Comparison between the bifurcation structure of homogeneous and heterogeneous 
networks using mean-field models.  The parameters are as in Table 1, except 
the applied current which is normally distributed with mean and variance as shown and 
$g_{syn}$ which varies as shown. Blue curve is equilibrium point, vertical black/blue
lines denote stable/unstable periodic orbits. Red dots denote bifurcation points.
(a) Homogeneous case. Numerical bifurcation analysis of MFI displays two subcritical 
Hopf bifurcations: one at a low $g_{syn}$ value and one at a high value.  
(b) Heterogeneous case.  Numerical bifurcation analysis of MFII also displays
two Hopf bifurcations, but the one at the low $g_{syn}$ value is supercritical.  
This makes bursting at low $g_{syn}$ values less robust in the heterogeneous case 
as discussed in the text.
} \label{fig5}
\end{figure}

In \cite{us2} we carried out a numerical bifurcation analysis for a homogeneous 
network. The mean-field equations in this case, which  are the same as MFI with 
$\meanIapp$ replaced by $I_{app}$, indicate that the transition from tonic
firing to bursting occurs via the following sequence of bifurcations. 
The stable bursting limit cycle is created in a saddle node bifurcation of 
(nonsmooth) limit cycles.  The smaller, unstable limit cycle becomes smooth 
in a grazing bifurcation and then disappears in a subcritical-Hopf bifurcation 
which destabilizes the equilibrium point corresponding to tonic firing.  This 
transition is shown in Figure \ref{fig5a} when $I$ is held fixed, and $g_{syn}$ 
is varied.    The bursting limit cycles are created at a low $g_{syn}$ value 
and destroyed at a high $g_{syn}$ value. The sequence of bifurcations is
the same in both cases. 

\begin{figure}
\centering
\subfigure[Hopf bifurcation from MFII and full network simulations with $\langle I\rangle=1500,\sigma_I=500$]{\includegraphics[scale=0.52]{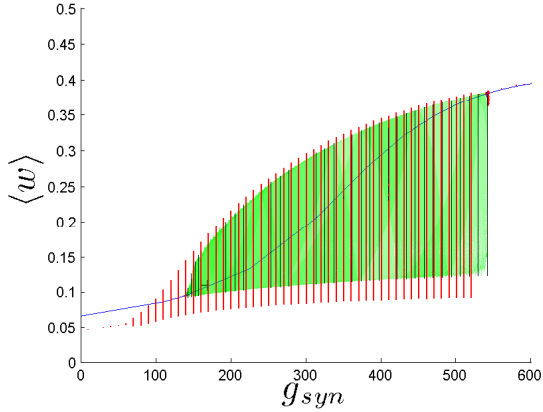} \label{fig6a}}
\qquad 
\subfigure[Hopf manifolds determined from mean-field models]{\includegraphics[scale=0.52]{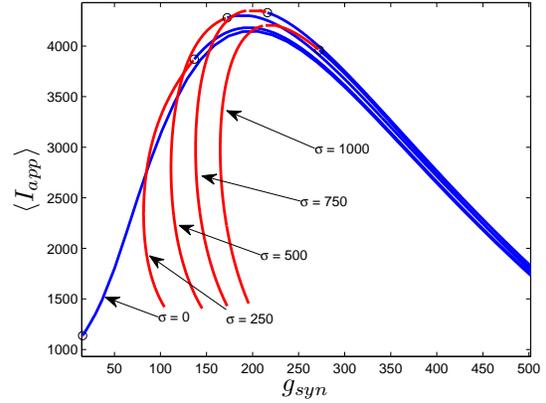} \label{fig6b}}
\\ 
\subfigure[Bursting regions from full network simulations]{\includegraphics[scale=0.5]{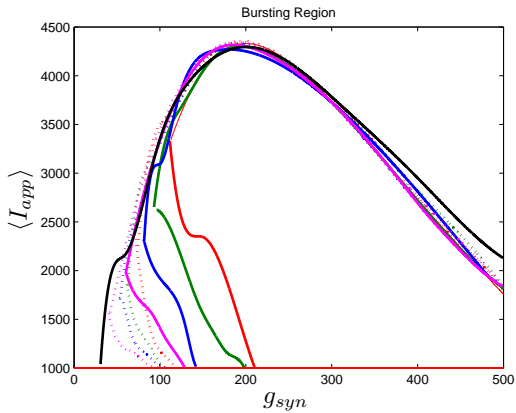} \label{fig6c}}
\qquad 
\subfigure[Bursting region from MFII and full network simulations with $\sigma_I=1000$]{\includegraphics[scale=0.52]{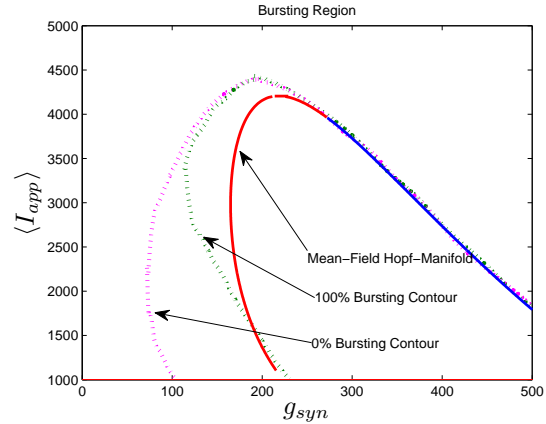}
\label{fig6d}}
        \caption{Comparison between numerical bifurcation analysis of MFII and direct 
simulation of the full network. The parameters are as in Table 1, except $g_{syn}$ 
varies as discussed below and the applied current which is normally distributed 
with mean and variance as discussed below. 
(a) Simulations of a network of 10,000 neurons with $\meanIapp$ and $\sigma_I$ as shown were run at discrete values of $g_{syn}$ for 2000 ms.  The last 400 (ms) of simulation time is plotted (in red), showing the stable limit cycle oscillation for different $g_{syn}$ values.  This is compared to numerical continuation of the MFII limit cycle and equilibrium (in green and blue).  Both the actual network and MFII appear to undergo a supercritical Hopf bifurcation for low $g$ values and a subcritical Hopf for high $g$ values. 
(b) The Hopf manifolds for the mean-field systems with $\sigma_I$ as shown.  Red denotes supercritical Hopf bifurcations, and blue denote subcritical Hopf bifurcations.  The black circles denote codimension 2 Bautin bifurcation points. 
(c) Simulations of a network of 1000 neurons run on a discrete mesh of $\meanIapp$ and $g_{syn}$ values.  The 0\% (dotted line) and 100\% (solid line) network bursting contours for $\sigma_I = 0, 250, 500, 750,$ and $1000$ pA are coloured in black, magenta, blue, green, and red, respectively.   The curves are spline fits to the actual contours.
(d) MFII Hopf manifold and spline fits to the 0\% bursting and 100\% bursting contours of the actual network for $\sigma_I=1000$. } \label{fig6}
\end{figure}

Using MFII, we numerically confirm that, as for the homogeneous network, the 
mean-field system of the heterogeneous network undergoes a Hopf bifurcation 
as the network transitions from tonic firing to bursting.  However, as shown in 
\ref{fig5b}, with $\meanIapp$ held fixed the transitions for low $g_{syn}$ and
high $g_{syn}$ are not the same. For high $g_{syn}$ the transition is the
same as the homogeneous case. For low $g_{syn}$ the transition occurs via
the following sequence of bifurcations.  A supercritical Hopf bifurcation 
destabilizes the equilibrium point corresponding to tonic firing and creates 
a stable limit cycle.  This limit cycle is smooth and hence corresponds not 
to bursting, but to firing with an oscillatory firing rate. 
This limit cycle then grows until it grazes
the switching manifold and becomes a non-smooth, bursting limit
cycle in a grazing bifurcation  (in Figure \ref{fig5b} this occurs at 
$g_{syn}\approx 150$).  We verified this prediction of the mean-field
model by running direct simulations of a network of 10,000 neurons with fixed 
$\meanIapp$, $\sigma_I$ while varying the $g_{syn}$ value.  As shown
in Figure~\ref{fig6a}, when the steady state mean variables are plotted vs $g_{syn}$, 
the supercritical nature of the Hopf bifurcation in the large network is 
apparent. 

To further investigate the heterogeneous case, we used MATCONT to carry out two 
parameter continuation of the Hopf bifurcation for MFII with four different values 
for the standard deviation of $I_{app}$: $\sigma_I = 250,500,750,1000$ pA.
As shown in Figure \ref{fig6b}, in all cases
there appears to be a codimension-2 Bautin (or generalized Hopf) bifurcation on 
the Hopf manifold, with the Hopf being supercritical on the left boundary before this point and subcritical after.  By contrast, in the homogeneous case ($\sigma_I=0$ line in
Figure \ref{fig6b}) the bifurcation is subcritical everywhere on the Hopf manifold.

Further verification of the mean-field results can be found in the direct numerical
simulations of a network of 500 neurons shown in Figure~\ref{fig6c}.
The simulations were run on a $50\times50$ mesh in the $g_{syn}$ vs $\meanIapp$ parameter space, using five different
values for the standard deviation of $I_{app}$: $\sigma_I = 0, 250,500,750,1000$ pA.  Note that $\sigma_I=0$ is the homogeneous network.  The proportion of bursting neurons, $p_{burst}$, was computed using equation (\ref{pbursteq}) (see Appendix A).  The 0\% and 100\% bursting contours can be seen in \ref{fig6c} and \ref{fig6d}.   In these figures, there appear to be two kinds of transitions from tonic firing to bursting.  Along the (lower) left part of the boundary of the bursting region, the transition is gradual:  the proportion of bursting neurons gradually increases from $0$ to $100\%$.  Along the rest of the boundary, however, the whole network transitions to bursting simultaneously.  This agrees with the prediction from the mean-field model that two different bifurcations occur along the bursting boundary.  Note also that the size of the entire bursting region and the 100\% network bursting region get smaller as the level of heterogeneity ($\sigma_I$) increases. 

Let us reiterate the primary differences between supercritical and subcritical Hopf 
induced bursting seen in Figures \ref{fig5} and \ref{fig6}. 
First, the subcritical case allows for bursting a lower $g_{syn}$ values than the 
supercritical case. This is because in the subcritical case bursting is initiated via a
a saddle-node of limit cycles bifurcation which occurs to the {\em left} of the Hopf 
bifurcation, while in the supercritical case, bursting starts to the {\em right} of the Hopf
in a grazing bifurcation. Second, the transition to bursting is sharp in the subcritical
case and gradual in the supercritical case. 
The supercritical Hopf bifurcation is consistent with the gradual transition from bursting to firing seen in Figures~\ref{fig6c} and \ref{fig6d}. When only a few neurons are bursting and the rest have oscillatory firing rates the corresponding mean behavior is a limit cycle with small amplitude. As more and more neurons become bursting this increases the amplitude of the limit cycle of the mean behavior until it grazes the switching manifold. In the subcritical case, the
saddle-node of limit cycles involves large amplitude limit cycles, corresponding to all the neurons being in the bursting state.

The bifurcation manifolds in Figure \ref{fig6b}  are both qualitatively 
and quantitatively accurate descriptions of the behavior of the actual network.  For example, in the actual network simulation, the bursting region decreases as $\sigma_I$ increases (see Figure \ref{fig6c}). This same behavior is displayed by the MFII equations, albeit to a greater degree, as shown in Figure \ref{fig6b}.  
However, there is a greater degree of quantitative error for lower values of $g_{syn}$ and larger values of $\sigma_I$.  In particular, for a fixed value
of $\sigma_I$ MFII predicts that the Hopf bifurcation occurs at a higher 
value of $g_{syn}$ than occurs in the real network (compare Figures \ref{fig6b} and 
\ref{fig6c}, and \ref{fig6d} directly) and this prediction error seems to increase as $\sigma_I$ increases.
This is why MFII indicates the network should be tonically firing when 
$\sigma_I$ is high (in Figure \ref{fig2d}, for example).

Taken together, these results indicate that for small $g_{syn}$, network induced bursting via adaptation is not robust to heterogeneity in the applied current.  This occurs for qualitative and quantitative reasons, both related to the Hopf bifurcation associated with the left boundary of the bursting region.  Qualitatively, the addition of heterogeneity causes this bifurcation to change from subcritical to supercritical making the bursting less robust for small $g_{syn}$ values.  Quantitatively, the $g_{syn}$ value of this bifurcation increases when the heterogeneity becomes stronger, while the value of the Hopf bifurcation associated with the right boundary does not change appreciably. Thus the size of the bursting region decreases with increasing heterogeneity.

\subsubsection{Bifurcation Types and Manifolds Using MFIII}\label{mf3sec}

It is difficult to use MATCONT with MFIII as MFIII is an infinite 
dimensional dynamical system, as it is a PDE. 
However, the existence of equilibrium points can be determined using 
standard root finding algorithms.   While direct bifurcation analysis is difficult to implement in this situation, one can use properties of the firing rate to describe, qualitatively and quantitatively, any transitions between network states.  This will be the approach
of this section. To begin, we consider networks that are tonically firing, we then proceed to the study of bursting networks.  

For a network of neurons with heterogeneity in the parameters, even if all the neurons are tonically firing, one cannot find a steady state firing rate for the network, as in the case of a homogeneous network.  The parameter heterogeneity creates a distribution of steady state firing rates across the network.  While the mean-field equations by themselves can only determine the mean of this distribution, with an added assumption we can approximate the distribution of steady state firing rates for the network with a great degree of accuracy. 

Consider a network with just one heterogeneous parameter, $\beta$. 
Assume that the steady state firing rate of each neuron in the network 
can be related to its value for the heterogeneous parameter: $R_i=g(\beta)$. 
Assume further that one can approximate this function by
the steady state value of $\langle R_i(t)|\beta\rangle$:  
\begin{equation}
g(\beta) \approx \langle R_i|\beta \rangle. 
\end{equation}
This is easily determined through direct simulation of MFIII, \eqref{wone3}-\eqref{MF3R}, until the system reaches steady state.  Treating $g$ as the 
transformation of a random variable,
one can determine the steady state distribution of firing rates in the network, $\rho_R(r)$, through the standard theorem on transforming random variables:
\begin{equation}
\rho_R(r) = \rho_\beta(g^{-1}(r))\left|\frac{d}{dr}g^{-1}(r)\right|.
\label{invmagic}
\end{equation}
which can be found in any standard textbook on probability theory (such as \cite{STATS1}).  Note that we must assume that $\langle R_i | \beta\rangle$ is monotonic and invertible for this procedure to be valid.

We carried out this computation for a network of 1000 neurons with a
normal distribution in either $I$, $g$, or $w_{jump}$. Details of the implementation
can be found in Appendix D We numerically determined the distribution of 
steady state firing rates for the neurons in the full network through 
\begin{equation}
R_i = \frac{1}{ISI_{i,last}}, \quad i = 1,2,\ldots N
\end{equation}
where $ISI_{i,last}$ is the last interspike interval for the $i^{th}$ neuron 
measured from a lengthy (1000 ms) simulation.  
Figure \ref{fig7} shows the results of the two approaches. The blue curve in 
the left column shows the distribution of parameter values. This is used
in MFIII to calculate the predicted distribution of firing rates, which is 
the dashed red curve in the right column. The solid blue curve in the right 
column is
the computed distribution of firing rates from numerical simulation of 
the full network equations. There is excellent agreement between the
firing rate distributions in all cases.

We carried out the same computations with a bi-modal distribution 
in $I$, $g$, or $w_{jump}$, generated by mixing normal unimodal distributions  
(see Appendix C). This is one way of representing a network with two 
subpopulations of neurons with different parameters. The mean field approach 
again gives an excellent approximation to the qualitative and quantitative properties of the steady state distribution of firing rates, as shown in the right column of Figure \ref{fig8}.  

The above approach is only valid when the network is tonically 
firing. In this situation the steady firing rates of the network and the
individual neurons are constant. When the network  leaves the tonic firing 
regime, however, these steady state firing rates become oscillatory.  In
the case of bursting, the amplitude of the oscillation is large enough that
the firing rate goes to zero for intervals of time. Whether or not the neurons
are bursting, oscillatory firing rates cannot be represented as a simple distribution of firing rates. However, with some additional work we can use the 
tools developed above to determine what proportion of neurons in the network is bursting.  This is a statistical alternative to direct bifurcation analysis.

From simulations of the full network, we know that when the variance in the heterogeneity 
is large enough, not all of the neurons necessarily display the behavior predicted from the mean-field equations. For example, Figure~\ref{fig9} shows
simulations of a network where the mean-field equations display an oscillatory 
firing rate which does not quite go to zero.  The spike time raster plot of the
full network (Figure~\ref{fig9a}) shows that some neurons {\em are} bursting while 
others are tonically firing with an oscillatory firing rate. However, simulation
of the corresponding mean-field equations (Figure~\ref{fig9b}, dashed line) reveals 
only an oscillation, not bursting. While this is consistent with the behaviour of the 
network mean variables (Figure~\ref{fig9b}, solid line), we have lost the information that
some of the neurons are bursting.  Similarly, one can
find examples where the mean-field equations exhibit bursting, but
not all neurons in the network are bursting. Thus, it would be useful to have
more information about individual neuron behaviour.  MFIII can be used to obtain such information.
s.   

In the situation described above, the steady state mean network firing
rate is oscillatory. We will denote this oscillatory solution as $\gamma$.  We interpret $\gamma$ to be the limit cycle parameterized by $\gamma(t)$ with $ (s(\gamma(t)),\langle w(\gamma(t)) | \beta \rangle)$ being the graph of the limit cycle in phase space.  We will denote the period of the limit cycle as $T$.  
In this case, ``steady state" firing rate for neuron $i$ will be a periodic
function of time: $\bar R_i(t)$, which depends on $\gamma$  and the value of 
the parameter $\beta$ associated with the neuron:
$\bar R_i(t) = g(\beta,\gamma(t))$,
for $t\in[0,T]$.    To proceed, we make the same assumption as above, that 
\begin{equation}
g(\beta,\gamma(t)) \approx \langle R_i(\gamma(t)) | \beta \rangle 
\end{equation}
where $ \langle R_i(\gamma(t)) | \beta \rangle $ is the oscillatory firing rate associated with the steady state limit cycle $\gamma$ in MFIII.  
An example of the graph of the steady state limit cycle derived from MFIII 
is shown in Figure \ref{fig9c}. In this visualization we can clearly see that
part of the network is bursting (blue) while the rest is tonically firing with an
oscillatory firing rate (green).  Integration of this limit cycle over the heterogeneous
parameter returns the ``mean" limit cycle (Figure~\ref{fig9d}).

We now use this setup to approximate $p_{burst}$, the proportion of neurons
in the network that are bursting during the network level oscillation $\gamma$.
Noting that $\langle R_i(\gamma(t)|\beta)\rangle\ge 0$, for all $t\in [0,T]$, 
the tonically firing neurons correspond to those $\beta$ values for which 
$ \langle R_i(\gamma(t)) | \beta \rangle > 0$, for all $t\in [0,T]$.  
Thus we define, $p_{tonic}$, the proportion of tonically firing neurons 
in the network via
\begin{equation}
p_{tonic} = \int_\beta X\left(\left[ \min_{t\in[0,T]} \langle R_i(\gamma(t)) | \beta \rangle \right] >0\right) \rho_\beta(\beta) d\beta
\end{equation}
where $X$ is the usual indicator function.  Similarly, the proportion of 
quiescent (nonfiring) neurons is given by 
\begin{equation}
p_{q} =  \int_\beta X\left(\left[ \max_{t\in[0,T]} \langle R_i(\gamma(t)) | \beta \rangle \right] =0\right)  \rho_\beta(\beta) d\beta.
\end{equation}
Recall that the bursting neurons correspond to those $\beta$ values such that
$ \langle R(\gamma(t)) | \beta \rangle = 0$, for some subinterval of $[0,T]$.  
Thus we must have
\begin{equation}
p_{burst} = 1 - p_{q} - p_{tonic}.
\end{equation}

We compute these values as follows.  First we numerically integrated MFIII 
until the steady state oscillation $\gamma$ is reached.  This is an oscillation of $\langle w(t) | \beta \rangle$, and $s(t)$.  The corresponding oscillatory
firing rate $\langle R_i (\gamma(t))| \beta\rangle$ is computed through 
equation \eqref{MF3R} as a function of $\beta$ on the limit cycle $\gamma$.  
One can then determine
\begin{eqnarray}
m(\beta) &=& \min_{t\in[0,T]}  \langle R(\gamma(t)) | \beta \rangle,\\
M(\beta) &=& \max_{t\in[0,T]}  \langle R(\gamma(t)) | \beta \rangle,
\end{eqnarray}
and the integrals simplify to 
\begin{eqnarray}
p_{tonic} = \int_\beta H(m(\beta))\rho_\beta(\beta)\,d\beta \\
p_{q} = \int_\beta H(-M(\beta))\rho_\beta(\beta)\,d\beta
\end{eqnarray}
where $H$ is the Heaviside function, with $H(0) = 1$.   

Numerical results for a network with single heterogeneous parameter are shown in Figure \ref{fig10}.  For Figure \ref{fig10a}, the proportion of bursting neurons was computed, using the method described above, at each point in a mesh on the $g_{syn}$ vs. $\meanIapp$ parameter space. This data was then used to  generate the $p_{burst}$ contours.  For Figure
\ref{fig10b} the actual network was simulated to steady state at each point of a slightly
coarser mesh. The proportion of bursting neurons at each point was computed according to 
equation 
\eqref{pbursteq} in Appendix B and used to generate the $p_{burst}$ contours.  The results of the mean-field computation are both qualitatively and quantitatively accurate.  In particular,
MFIII recovers the gradual transition to bursting on the left boundary of the bursting region and the abrupt transition to bursting on the right. It should be noted that it is much faster, by approximately an order of magnitude, to run a mesh of integrations over MFIII then it is to run mesh over an actual network.

\subsubsection{Inverting a Steady State Firing Distribution to Determine the Distribution of Parameters Using MFIII}\label{invsec} 

Many parameters for neuron models are difficult to measure directly using electrophysiology.   However, a distribution of firing rates across a network of neurons is relatively easy to measure using intracellular recordings, or can be estimated using measurements from multi-electrode recordings and spike sorting algorithms, among other methods \cite{Buzsaki04,Grewe}.  We have seen in the previous section that, given a distribution of heterogeneities, MFIII can predict the steady state distribution of firing rates.  Here we show that one can invert this process to yield a distribution of parameters given a steady state distribution of firing rates.  

We assume that only the firing rate distribution is known, and denote it 
$\rho_R(r)$ as above. We then proceed as in the previous section, assuming 
that the steady state firing rate for a particular neuron is some function of the heterogeneous parameters $R_i = g(\beta)$  and that this function is well approximated by $\langle R_i |\beta \rangle$. 
Under these assumptions, one can solve for the distribution of parameters 
$\beta$ using
\begin{equation}
\rho_\beta(\beta) = \rho_R(g(\beta))\left| \frac{d}{d\beta}g(\beta)\right| \label{magic}
\end{equation}
which follows from standard statistical theorems on the transformations of random variables \cite{STATS1}.  
Note that we need to assume that $\langle R_i | \beta\rangle$ is differentiable
for this procedure to be valid. 

The primary problem we face in using this approach to approximate the distribution $\rho_\beta(\beta)$ is that we need to determine the steady state values of the  function $\langle R_i | \beta\rangle$.  However, a cursory look at the equations for MFIII shows that these in fact depend on $\rho_\beta(\beta)$, the function we are trying to find, through the equation for $s$: 
\begin{equation}
\dot{s}=  -\frac{s}{\tau_s} + s_{jump}\int_\beta\langle R_i(t)|\beta\rangle\rho_\beta(\beta)\,d\beta.
\end{equation}
Fortunately, however, this problem disappears when we look at the
steady state value for $s$: 
\begin{equation}
\bar s = \tau_s s_{jump} \int_\beta\langle R_i|\beta\rangle\rho_\beta(\beta)\,d\beta = \tau_s s_{jump} \langle R_i \rangle.
\end{equation}
Here $\langle R_i \rangle$ is the unconditioned steady state mean of the firing rate distribution.  This information is readily available, as we have assumed we know the steady state distribution, $\rho_R(r)$, and determining the first moment is numerically trivial. 

Putting the expression for $\bar s$ into the steady state equation for
$\langle w|\beta\rangle$ yields a set of coupled equations: 
\begin{eqnarray}
{\langle w | \beta \rangle} &=& \tau_w w_{jump} \langle R_i|\beta \rangle, \label{ss1}\\
\langle R_i|\beta \rangle &=&    \left\{     \begin{array}{lr}  \left[\int_{V}\frac{dv}{G_1(v,\tau_ss_{jump}\langle R \rangle,{\langle w|\beta\rangle},\beta)} \right]^{-1}& : H({\langle w|\beta \rangle},\bar{s},\beta)\geq 0\\      0 & H({\langle w|\beta \rangle},\bar{s},\beta)<0     \end{array}.   \right.\label{ss2} 
\end{eqnarray}
These may be solved for $\langle w|\beta\rangle$ and 
$\langle R_i | \beta \rangle$ by discretizing in $\beta$ and numerically 
solving the resulting system at each grid point with any standard root finding algorithm.  

Alternatively, one can set $s$ to its equilibrium value in MFIII
and numerically integrate the resulting equation:
\begin{eqnarray}
\langle w | \beta\rangle' &=& -a\langle w | \beta\rangle + w_{jump}\langle R_i(t)|\beta\rangle, \label{woneI} \\
\langle R_i(t)|\beta \rangle &=&    \left\{     \begin{array}{lr}  \left[\int_{V}\frac{dv}{G_1(v,\tau_ss_{jump}\langle R_i \rangle,\langle w|\beta\rangle,\beta)} \right]^{-1}& :H(\langle w|\beta \rangle,s,\beta)\geq 0 \\      0 &H(\langle w|\beta \rangle,s,\beta)<0    \end{array}   \right.\label{MF3I} 
\end{eqnarray}
until it reaches steady state, which will determine ${\langle w|\beta\rangle}$ and
$\langle R_i|\beta\rangle$.  Note that this approach will only work if the tonic firing equilibrium of the original mean-field system MFIII is asymptotically stable.

\begin{figure}
\centering
\subfigure[Distribution of $I_{app}$ with $\meanIapp = 4500$ pA, $\sigma_I = 1000$ pA]{\includegraphics[scale=0.55]{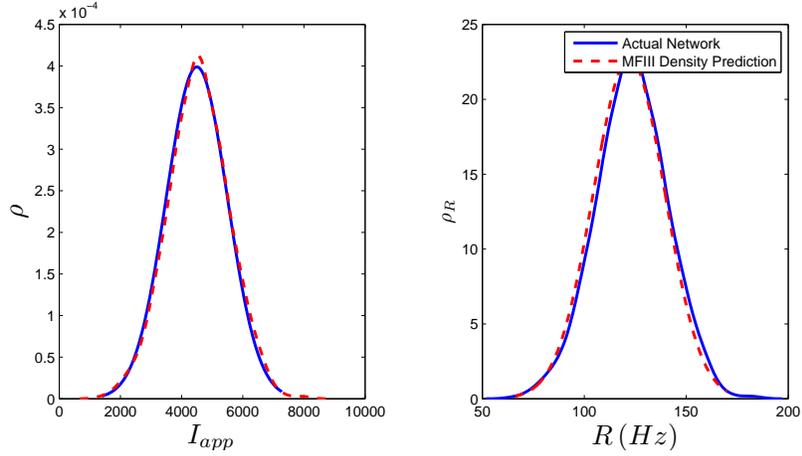}
\label{fig7a}}
\\ 
\subfigure[Distribution of $g_{syn}$ with $\meangsyn = 200$ nS, $\sigma_g$ = 50 nS]{\includegraphics[scale=0.55]{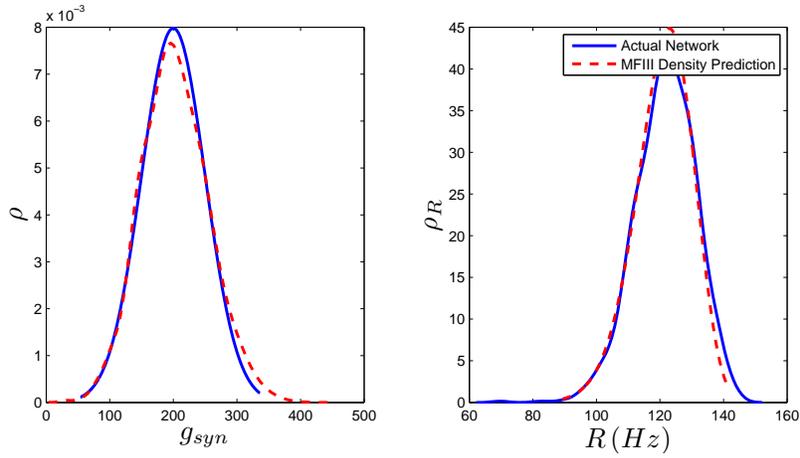}
\label{fig7b}}
\\
\subfigure[Distribution of $W_{jump}$ with  $\meanWj = 200$ pA, $\sigma_W$ = 50 nS]{\includegraphics[scale=0.55]{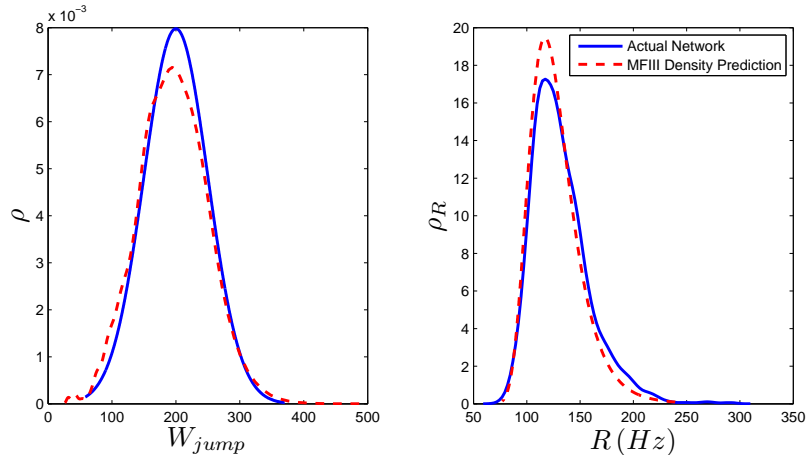}
\label{fig7c}}
        \caption{Heterogeneity in $I_{app}$, $g_{syn}$, or $W_{jump}$ leads to heterogeneity in the firing rate. As described in section~\ref{mf3sec}, given the parameter distribution 
(solid curves, left column) MFIII can be used to estimate the corresponding 
distribution of firing rates in the network (dashed curves, right column). 
As described in section~\ref{invsec}, given the the steady state firing rate distribution 
of an actual network (solid curves, right column) MFIII can be used to estimate
the parameter distribution in the network (dashed curves, left column).
The network firing rate distribution is estimated using a histrogram.  The calculations were carried out on a network of 1000 neurons. Parameters, other than those indicated, can be found in Table \ref{table1}. } \label{fig7}
\end{figure}

\begin{figure}
\centering
\subfigure[Distribution of $I_{app}$ with $\mu_1 = 4500$ pA, $\sigma_1 = 500$ pA, $\mu_2 = 7000$ pA, $\sigma_2 = 1000$ pA, $m= 0.7$]{\includegraphics[scale=0.52]{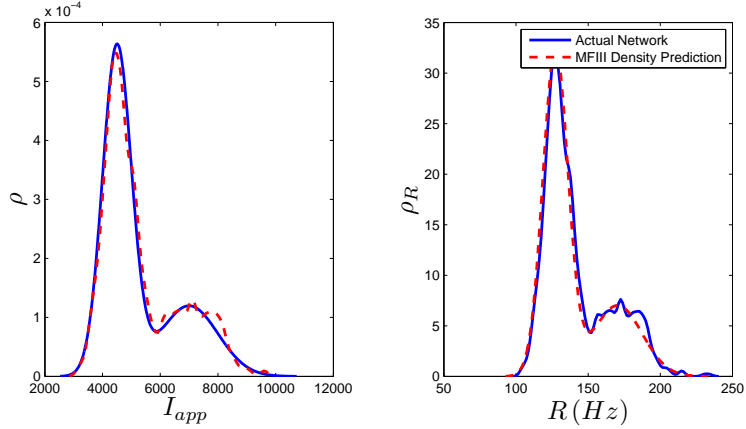}
\label{fig8a}}
\\ 
\subfigure[Distribution of $g_{syn}$ with $\mu_1 = 100$ nS, $\sigma_1 = 30$ nS, $\mu_2 = 300$ nS, $\sigma_2 = 50$ nS, $m= 0.4$]{\includegraphics[scale=0.52]{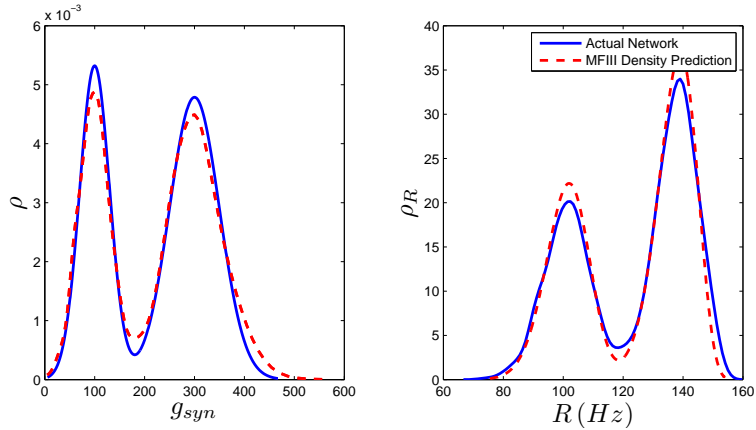}
\label{fig8b}}
\\
\subfigure[Distribution of $W_{jump}$ with $\mu_1 = 300$ pA, $\sigma_1 = 50$ pA, $\mu_2 = 75$ pA, $\sigma_2 = 20$ pA, $m= 0.6$]{\includegraphics[scale=0.52]{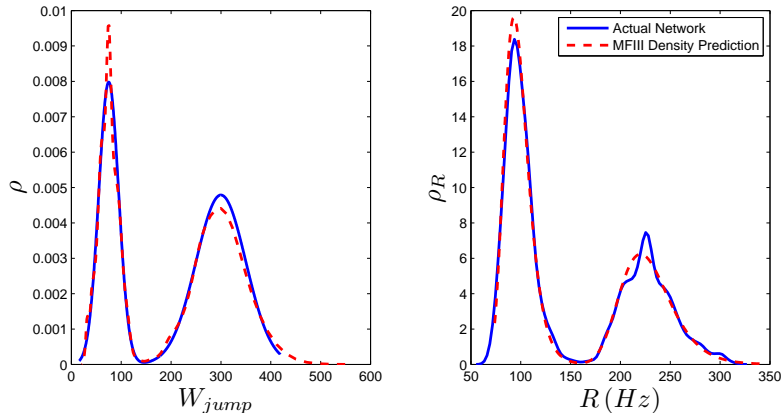}
\label{fig8c}}
       \caption{Bimodal distributions in $I_{app}$, $g_{syn}$, and $W_{jump}$ lead to bimodal distributions in the firing rate.  These bimodal parameter distributions are generated through distribution mixing of two normal subpopulations with standard deviations and means as indicated.  See Appendix B for details.  The distribution of the firing rate or the distribution of the parameter can be computed using MFIII if one knows the complementary distribution. See sections \ref{mf3sec} and \ref{invsec} for details.
Curve descriptions are as given in Figure~\ref{fig7}.  
Parameters, other than those indicated, can be found in Table \ref{table1}.  The calculations were carried out on a network of 1000 neurons.} \label{fig8}
\end{figure}

\begin{figure}
\subfigure[Spike time raster plot of network behaviour]{\includegraphics[scale=0.52]{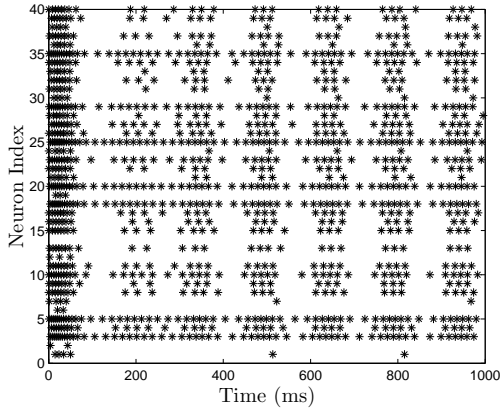}
\label{fig9a}}
\qquad 
\subfigure[Network and MFIII mean variables show oscillation]{\includegraphics[scale=0.52]{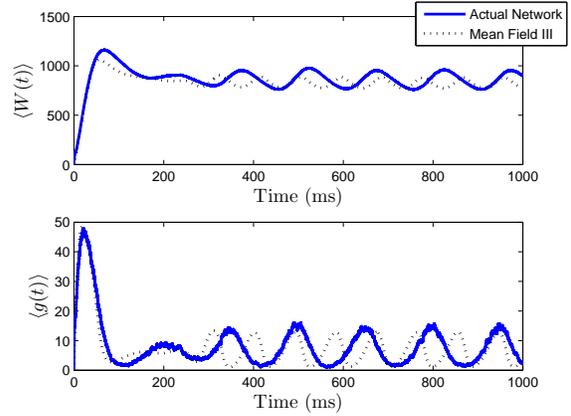}
\label{fig9b}}
\\
\subfigure[MFIII limit cycle in conditional variables]{\includegraphics[scale=0.29]{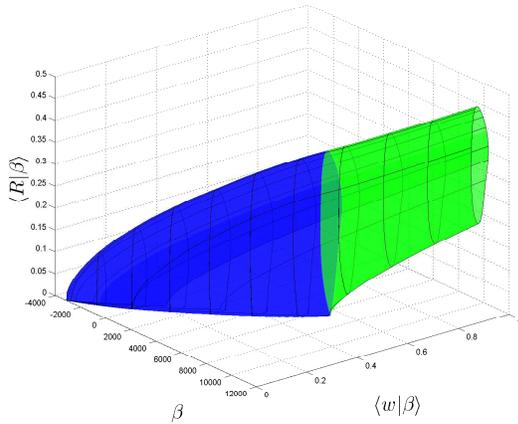}
\label{fig9c}}
\qquad 
\subfigure[MFIII limit cycle in mean variables]{\includegraphics[scale=0.52]{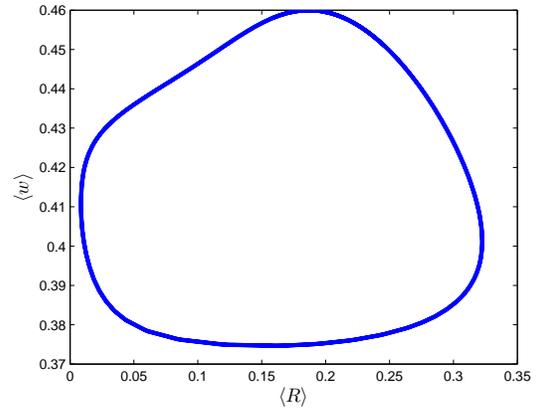}
\label{fig9d}}
\caption{Visualizing a limit cycle in a heterogeneous network. Numerical simulation of MFIII and a network of 1000 neurons with heterogeneity in the applied current.  Parameters are as given in Table \ref{table1} except $g_{syn}=200$ nS, $\meanIapp = 1000$ pA   
and $\sigma_I = 4400$ pA. 
(a) Raster plot for 40 randomly selected neurons of the network.  Some of the neurons are bursting, while others are tonically firing, albeit with an oscillatory firing rate.  
(b) In the mean variables, the steady state behaviour of both the network and MFIII is an oscillation. 
(c) As MFIII is a partial differential equation, the steady state ``limit cycle" is actually a manifold of limit cycles, foliated by the heterogeneous parameter $\beta=I_{app}$.  Part of the manifold has cycles with $\langle R|\beta\rangle=0$ for an extended period of time (in blue), and the other part contains limit cycles that have $\langle R | \beta\rangle\ne 0$ during the entire oscillation.  We can classify neurons with the parameter values in blue as bursting, and those in green as oscillatory firing.  
(d) Averaging the limit cycle in (c) with respect to $\beta$ yields the mean limit cycle. }\label{fig9}
\end{figure}

\begin{figure}
\centering
\subfigure[MFIII, $\sigma_I = 500$ pA ]{\includegraphics[scale=0.80]{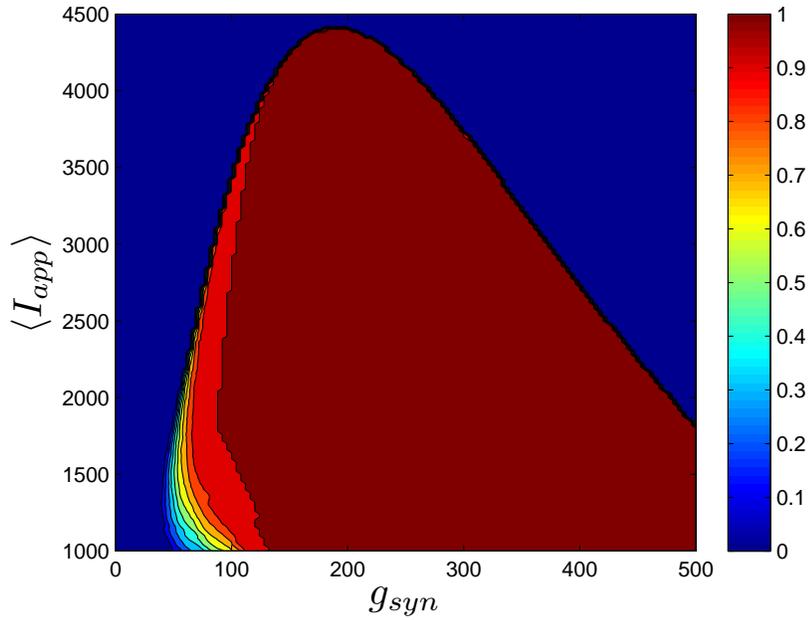}
\label{fig10a}}
\subfigure[Network, $\sigma_I = 500$ pA]{\includegraphics[scale=0.80]{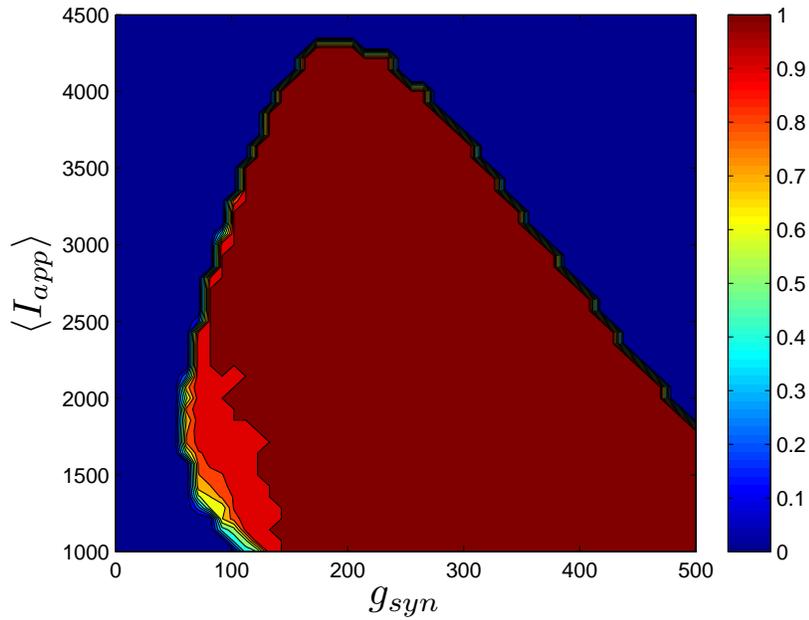}
\label{fig10b}}
\caption{The proportion of bursting neurons, $p_{burst}$ for MFIII and an actual network.  
(a) Using the techniques outlined in the text, MFIII is used to compute the proportion of bursting neurons, $p_{burst}$ at each point in a mesh over the parameter space. (b) Numerical simulations of a network of 500 neurons is used to compute the proportion of bursting neurons, $p_{burst}$.  All the parameters for both the MFIII system, and the actual network are
identical (see Table~\ref{table1}), except that MFIII is run over a finer mesh.  The results using MFIII are both qualitatively and quantitatively accurate.} \label{fig10}
\end{figure}

We have implemented this approach as follows.  A network of 1000 neurons is 
numerically integrated until it reaches its steady state firing rate.  The
distribution of firing rates over the network is found as described 
in the previous section.  The density function for this distribution, $\rho_R(r)$, 
is then estimated using the firing rate histogram.  Equations \eqref{woneI}-\eqref{MF3I} 
are numerically integrated until they reach steady state. We then substitute 
the estimate of $\rho_R(r)$ and the approximation 
$\langle R_i|\beta\rangle$ of $g(\beta)$ into (\ref{magic}) to determine the 
parameter distribution $\rho_\beta(\beta)$. See Appendix D for more details. 
Our results for unimodal and bimodal distributions are shown in 
Figure~\ref{fig7}
and Figure~\ref{fig8}, respectively. In the right column of each figure, the 
solid blue curve is the distribution of steady state firing rates from 
integration of the full network.  In the left column of each figure the dashed 
red curve is the estimate of $\rho_\beta(\beta)$ found using the procedure above, while the blue curve is the actual parameter distribution used in the network simulation.  We note that no information about the distribution of parameters is known in the estimation procedure, yet the numerical results are very accurate in both the unimodal (Figure \ref{fig7}) and the bi-modal case (Figure \ref{fig8}).  

Perhaps most interesting is that we can extend this technique to estimate the individual neuron parameter values, $\beta_i,\ i=1,\ldots,N$.  This again follows from the assumption that $g(\beta) = \langle R_i | \beta\rangle$ is the function that transforms the random variables $\beta_i$ into $R_i$.  If the function is invertible, then we can compute the individual $\beta_i$ through numerically inverting the steady state $\langle R_i | \beta\rangle$.  For example, when this technique is applied to a network where the only source of heterogeneity is $I$, the mean relative absolute error in the predicted values $\hat I_i$ versus the actual values $I_i$ is only 0.6\%.   The details about how to numerically invert for the individual parameter values are included in Appendix D.

While network level inversion of a single heterogeneous parameter is an important step forward, this is performed under very strong assumptions.  In particular, when performing this inversion, all of the heterogeneity in the firing rates is assumed to come from a single parameter.  Additionally, all the other parameters are assumed to be known.  These two assumptions are exceptionally strong and one has to take great care in inverting actual recorded firing rates from neurons that they be reasonably satisfied.

\subsection{Mean-Field Applications with Multiple Sources of Heterogeneity} 
In order for the mean-field applications to be useful for realistic neuronal networks, one needs to consider heterogeneity in more than one parameter. Recall
that the mean field systems derived in section~\ref{MFTsec} are valid for 
multiple heterogeneous parameters, one simply considers $\beta$ to be a 
vector instead of scalar. This presents some difficulties in implementation 
which we discuss in the section. The examples we consider will have
2 or 3 sources of heterogeneity, primarily in the parameters $I$, $g$ and $d$.  

Recall that MFII is given by the equations \eqref{wwone}-\eqref{vm2}. 
The main difficulty in dealing with MFII lies with the integral terms, 
which are now multiple integrals. For example:
\begin{equation}
\langle R_i \rangle = \int_{\beta_1}\int_{\beta_2}\ldots \int_{\beta_p} \langle R_i | \bm\beta \rangle \rho_{\bm \beta}(\beta_1,\beta_2,\ldots \beta_p) d\beta_p\ldots d\beta_2\beta_1 \label{BAMF}
\end{equation}
where $p$ is the number of heterogeneous parameters.  
In order to numerically integrate or carry out bifurcation analysis on 
MFII, these multiple integrals must be evaluated.  We have found that this is most easily done using a Monte-Carlo numerical integration scheme.   
Once this is implemented, bifurcation diagrams can be generated exactly
as for the case of one parameter heterogeneity: the equilibrium points
and smooth limit cycles are continued using MATCONT, while the nonsmooth
limit cycles are generated using numerical simulations.  

The integral term in MFIII can be dealt with in a similar way as to
that for MFII. Once this is implemented, the steady states and network properties
can be determined as
described in section 3.2, while numerical simulations can be used to follow stable periodic solutions.  We will use this approach later on in our case study on adaptation induced bursting.

\begin{figure}
\centering
\subfigure[Parameter distributions]{\includegraphics[scale=0.55]{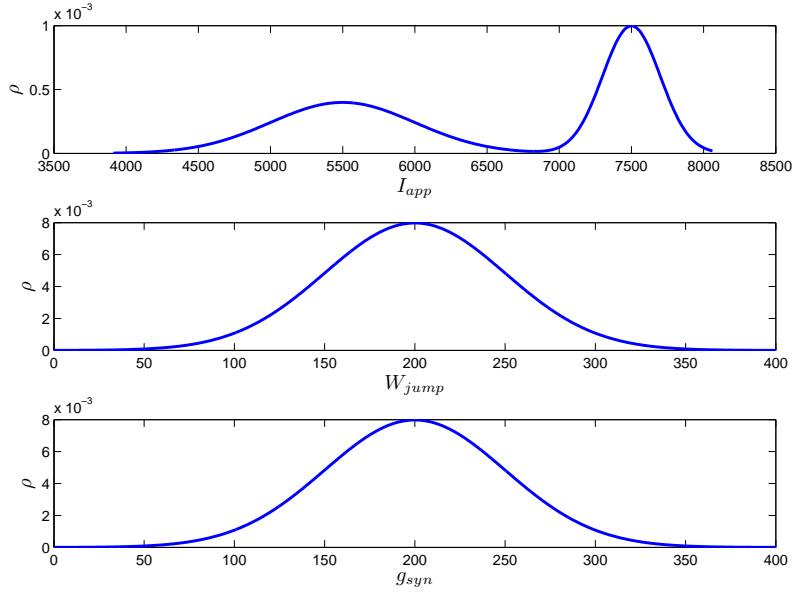}
\label{fig11a}}
\\ 
\subfigure[Resulting firing rate distributions]{\includegraphics[scale=0.55]{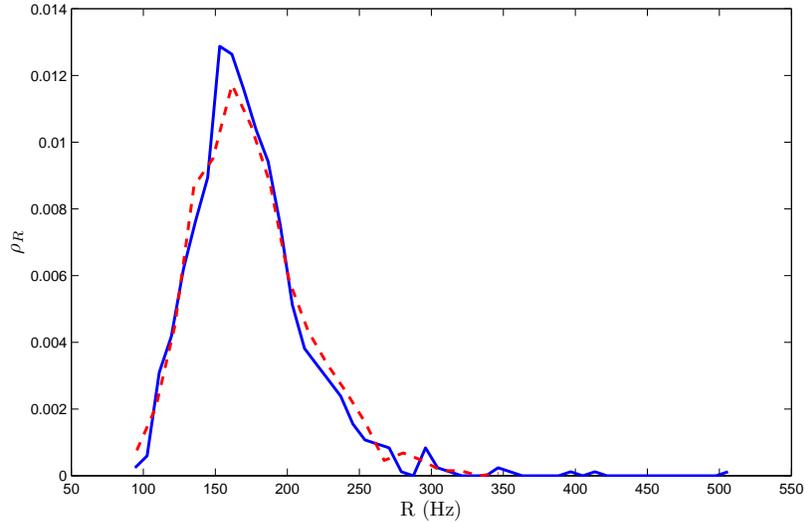}
\label{fig11b}}
\caption{A bimodal distribution in $I_{app}$ together with unimodal distributions in $g_{syn}$ and $W_{jump}$ as shown in (a) yields a unimodal distribution in the firing rate (dashed curve in (b)). The mean-field equations give a good estimate of this
distribution (solid curve in (b)). Simulations are for network of 1000 neurons.  
parameter given in Table~\ref{table1} except $I$, $g$ and $W_{jump}$.  These are distributed with $\sigma_w = 50$, $\langle W_{jump}\rangle = 200$, $\sigma_g = 50$, $\langle g_{syn}\rangle = 200$, $m=0.5$, $\langle I_{app,1} \rangle = 7500$, $\sigma_{I,1} = 200$, $\langle I_{app,2} \rangle = 5500$,  $\sigma_{I,2} = 500$.  Other combinations of bimodal and unimodal parameter distributions may yield bimodal firing rate distributions.  Note that this is different than the situation shown
in Figure~\ref{fig8}, where a single bimodal source of heterogeneity yielded a bimodal firing rate distribution. 
} \label{fig11}
\end{figure}

Mean-field III can also be used to determine steady state firing rates following
the procedure in section \ref{mf3sec}, however, one now has to discretize the 
equations over a multi-dimensional mesh.  While this approach is feasible, 
we found it is more efficient to predict the steady state firing rates of the 
individual neurons through the following interpolation scheme.  Given 
knowledge of the parameter distribution, we generate sample points 
$\bm\beta_i$ from this distribution.  
We then generate a steady state firing rate for each sample point, 
$R_i=g(\bm\beta_i)$, where $g$ is determined from MFIII as described in section
\ref{mf3sec}.  We interpolate over the $(\bm\beta_i,R_i)$ ordered pairs to determine the firing rates of the individual neurons, given knowledge of their parameter values.  If we only need the distribution of the 
firing rates, then the distribution of the $R_i$ is an estimate of this, 
without need for interpolation.  This approach has been applied to a network
of 1000 Izhikevich neurons with three simultaneous sources of heterogeneity, as 
shown in Figure \ref{fig11}.

The one application we found difficult to extend to the case of multiple
sources of heterogeneity was the mapping of the distribution of steady state 
firing rates to the distribution of parameters. There is a fundamental 
difficulty with this inversion problem: the firing rate distribution is 
one-dimensional, but the distribution of parameters is multidimensional.  
Thus, we leave further investigation of this problem for future work. 

\subsubsection{Bifurcation Analysis With Multiple Sources of Heterogeneity - Case Study}

To conclude our work, we consider a realistic model for a CA3 hippocampal
network of pyramidal cells. Hemond et al.  \cite{Hemond} classify CA3 
pyramidal cells into three types: weakly adapting, strongly adapting
and intrinsically bursting. We will focus on the effect on network bursting
of having two subpopulations: one strongly adapting and one weakly adapting. We 
use the Izhikevich model \eqref{IzV}-\eqref{Izhom_jump} with the parameters 
set up by \cite{us} (see Table~\ref{table1}), 
but include heterogeneity in $I_{app}$, $g_{syn}$ and the adaptation parameters $W_{jump}$, 
$\tau_{W}$.  The parameter distributions are generated through distribution mixing 
(see Appendix B) of normal distributions with the parameters given in Table~\ref{table2}.  
We have treated the mean values of $I_{app}$ and $g_{syn}$ from the strongly adapting 
subpopulation as the bifurcation parameters.  We also varied the proportion of strongly 
adapting neurons in the population, i.e., parameter $p$ in equation \eqref{mixedpdf}. 

\begin{figure}
\centering
\subfigure[Network bursting contours ]{\includegraphics[scale=0.70]{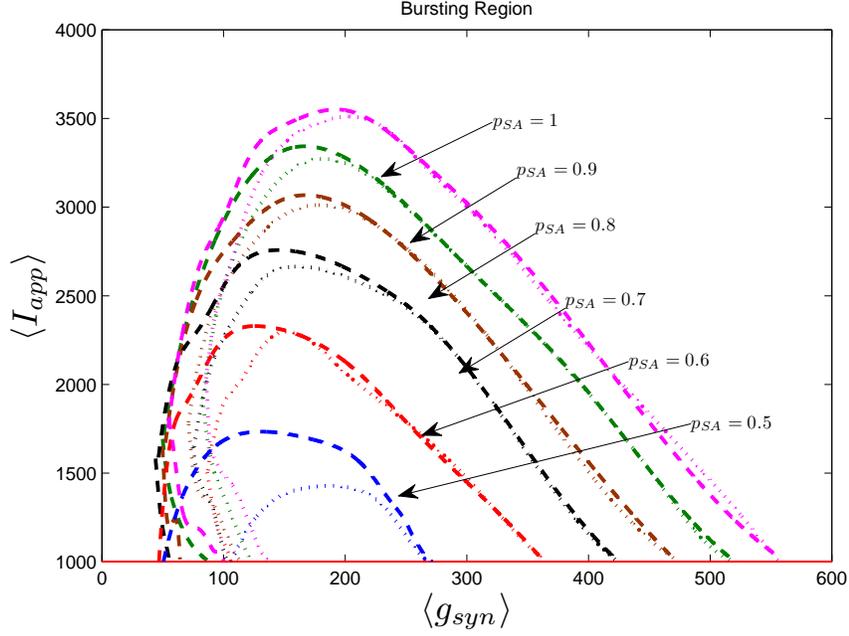}
\label{fig12a}}
\\
\subfigure[MFIII bursting contours]{\includegraphics[scale=0.70]{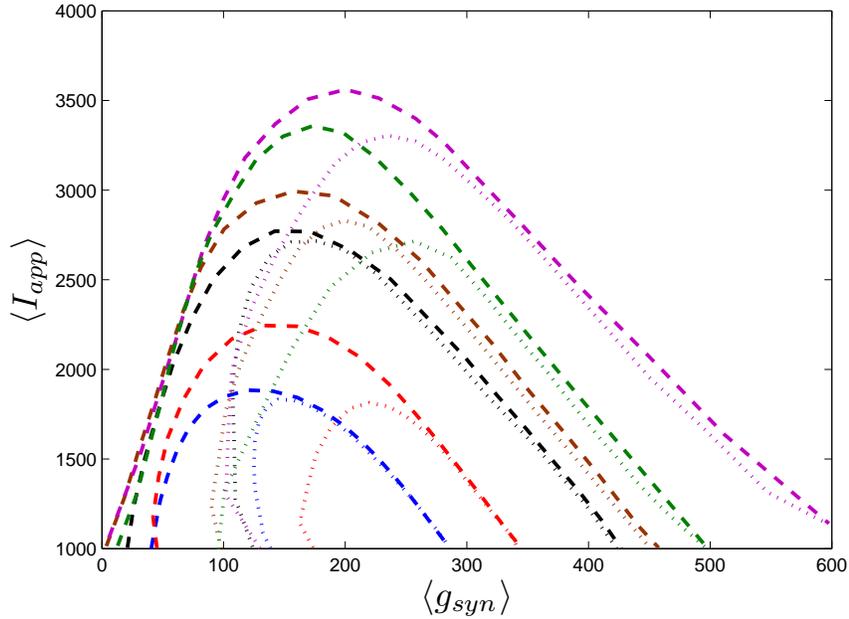}
\label{fig12b}}
\caption{Case study: the proportion of bursting neurons in a network with both strongly
adapting and weakly adapting neurons. Parameters are as in Table~\ref{table1} except
the $I_{app}$, $g_{syn}$, $W_{jump}$ and $\tau_W$ which have bimodal distributions generated
by distribution mixing with parameters given in Table~\ref{table2}. The parameter $p_{SA}$ 
represents the proportion of strongly adapting neurons in the network (see equation~\eqref{mixedpdf}).
The dashed line is the $0\%$ bursting contour, while the dotted line is the $100\%$ bursting contour.  As shown in both the network (a) and MFIII (b), the bursting regions becomes significantly smaller when the proportion of strongly adapting neurons decreases.  In all cases, the curves are smoother spline fits to the actual contours}\label{fig12}
\end{figure}

The 0\% and 100\% bursting contours for simulations over the two parameter mesh in the $\meanIapp,\meangsyn$ are shown in Figure \ref{fig12} for both the full network (Figure \ref{fig12a}) and MFIII (Figure \ref{fig12b}).   Numerical bifurcation analysis of MFII (not shown)
confirms that the bifurcations are similar to when $I_{app}$ was the only source of heterogeneity,
in particular, on the left boundary of the bursting region the Hopf bifurcation is supercritical, while on the right it is subcritical.  

As shown in Figure \ref{fig12}, when the proportion of strongly adapting neurons is decreased, the bursting region decreases.  However, unlike previous results \cite[Figure 10]{us2}, the decrease seems to be more pronounced in the high $g_{syn}$ region.   This is likely due to having truly heterogeneous distributions of parameters, as opposed to splitting a network into two different homogeneous subpopulations as was done in \cite{us2}.   
In all cases, it appears that heterogeneity shifts the bursting manifold 
outside of the low $g_{syn}$ region, which we suggest is the region of 
biologically plausible conductances \cite{us2}.

\begin{table}[htp]
\centering
\begin{tabular}{|c|c|c|}
\hline
Parameter & Strongly Adapting & Weakly Adapting\\
\hline
$\langle g_{syn} \rangle$ & 0-600 nS & 200 nS \\ 
\hline
$\langle \sigma_g \rangle$ & 0.5$\langle g_{syn} \rangle$ & 50 nS \\
\hline
$\langle I_{app} \rangle$ & 1000-4000 pA & 1200 pA \\
\hline
$\sigma_I$ & 500 pA & 500 pA \\ 
\hline 
$\langle W_{jump} \rangle$ &200 pA & 100 pA \\ 
\hline
$\sigma_{W_j}$ & 50 pA & 20 pA \\
\hline
$\langle \tau_w \rangle$ &200 ms & 50 ms \\
\hline 
$\sigma_{\tau_w}$ &50 ms & 10 ms \\
\hline
\end{tabular}
\caption{Table of parameters for the strongly and weakly adapting heterogeneous subpopulations}\label{table2}
\end{table}

\section{Discussion}

Building on the mean-field framework for networks of homogeneous oscillators, we extended the mean-field approach to networks of heterogeneous oscillators.   This was accomplished through the derivation of three separate mean-field systems, MFI, MFII and MFIII, with differing applications and regions of validity.  
We successfully applied numerical bifurcation analysis to MFI and MFII to 
aid in the understanding of the different behaviors that heterogeneous networks can display, and how they transition between these different types of behaviors.  
More importantly however, we have surpassed the natural limitation of 
mean-field systems: that they can only provide information about the first 
moments.  With a few additional tools, we used MFIII to derive information 
about distributions of firing rates, and even parameters, given some basic knowledge.   

Other researchers \cite{HM,Vlad} have derived firing rate distributions for heterogeneous
networks, however these have been derived under differing assumptions.  For example, the heterogeneous mean field systems 
studied by Hansel and Mato (Equations (5.5)-(5.7) in \cite{HM}) have similar integral terms to our MFII however they are firing rate models.  Our models are current/conductance based models.  The difference between these two types of equations arises from which time scale is the fastest, that of the synaptic current, or the firing rate.  If the firing rate time scale is assumed to be the fastest, then a differential equation for the synaptic current can be obtained (as in our case).  If the time scale of the synaptic current is assumed to be the fastest, then one obtains firing rate equations, as in \cite{HM}.   The fact that these two different limits result in different kinds of equations were first highlighted in \cite{DA} (section 7.2).   Additionally, no adaptation is contained in the rate models in \cite{HM}.   Finally, it is likely that the firing rate models shown in \cite{HM} cannot display period doubling bifurcations as they have a similar structure to MFII, which misses out on the more complicated bifurcations of the actual network that MFIII can reproduce, due to its PDE nature.   

The model of Vladimirski 
et al. \cite{Vlad} 
is formulated in terms of an input-output relation for the synaptic conductance, so has 
a different structure than ours. It involves a distribution of the synaptic 
depression variable so has some aspects similar to our MFIII, however, no PDE governing
the evolution of this variable is derived.

Dur-e-Ahmad et al. \cite{us} studied adaptation induced bursting in a network of 
homogeneous Izhikevich neurons, with parameters determined from experimental data on 
CA3 pyramidal neurons. They showed that, if the adaptation is strong enough, 
network bursting occurs in large regions of the parameter space consisting
of the synaptic conductance, $g_{syn}$, and the applied current, $I_{app}$.
In \cite{us2} we showed that the transition from tonic firing to bursting involves a 
saddle-node bifurcation
of non-smooth limit cycles, followed by a grazing bifurcation and a subcritical Hopf bifurcation.
For fixed $I_{app}$ greater than rheobase but sufficiently small, there is one 
transition from tonic firing to bursting at a low $g_{syn}$ value and another
from bursting back to tonic firing at a higher $g_{syn}$ value. Thus the
bursting region is a closed semi-circular region in the $g_{syn}$, $I_{app}$ 
parameter space.  In \cite{us2} we showed 
that the size of this bursting region is reduced if the network is split into two 
homogeneous subnetworks, one strongly adapting and one weakly adapting.
Here, we used the tools we developed to investigate how this adaptation 
induced network bursting is affected by heterogeneity in the parameters.  Somewhat surprisingly, we have found that adaptation induced network bursting is not very robust to heterogeneity.  This has been confirmed by direct simulations of the full network, bifurcation analysis using MFII and analysis of the proportion of bursting neurons in the network using MFIII.   This lack of robustness is caused by two changes to the homogeneous case:
\begin{enumerate}
\item  The low $g_{syn}$ Hopf bifurcation point moves towards higher values, 
thereby decreasing the size of the bursting region  
\item  The low $g_{syn}$ Hopf bifurcation switches from subcritical to supercritical. 
This has two effects:
\begin{itemize}
\item the bifurcation direction changes, eliminating the bursting at conductance values
less than the bifurcation value  
\item the initial limit cycles created by the bifurcation are small amplitude oscillations in the firing rate as opposed to full bursts, thus the transition to bursting moves to 
even higher conductance values. 
\end{itemize}
\end{enumerate}
Further, in networks with both weakly and strongly adapting neurons, heterogeneity
caused the high $g_{syn}$ Hopf bifurcation value to decrease when the
proportion of strongly adapting neurons is reduced.

Let us now put our results in the context of experimental results on the CA3
region.  Bursting is often seen in these studies \cite[Section 5.3.5]{hippobook}.
When the neurons have their synaptic inputs blocked however, the majority ($\approx 80$\%) of these pyramidal neurons do not display bursting, but different degrees of spike frequency adaptation \cite{Hemond}.  Thus, it would seem that 
adaptation induced network bursting should play a role in the CA3 network.  
However, the biophysically important part of the parameter region is in the low $g_{syn}$ region \cite{us2}. When this fact is taken in conjunction with our
results described above, this would seem to weaken the case that adaptation induced 
network bursting is the only source of bursting in CA3 networks. Some other
mechanism seems necessary. 

 In their study of hippocampal CA3 pyramidal neurons, Hemond and colleagues note that roughly 20\% of pyramidal neurons were intrinsically bursting.  That is, the neurons burst without any synaptic input for some input current.  It may by possible that a small subpopulation of intrinsically bursting neurons can facilitate bursting in the rest of the network, however this would depend on the conductance values connecting this particular subpopulation to the rest of the network.  This hypothesis can be tested relatively easily using a mean-field approach.  All that is required is to fit a two-dimensional adapting model to the intrinsically bursting neurons.  This is feasible, as has been previously noted, all the two-dimensional adapting neurons can be turned into intrinsically bursting neurons by simple parameter changes \cite{Izhikevich}.  The conductance parameter connecting this subpopulation to the rest is best treated as a bifurcation parameter, with some estimate of the range in which it lies in from physiological data.  
 
While an intrinsically bursting subpopulation is the most promising avenue of study with regards to hippocampal bursting, synaptic depression has also been shown to induce bursting in oscillators that cannot otherwise display this behavior.  In a model of the developing chick spinal cord, \cite{Vlad} found that heterogeneity actually makes the bursting more robust, as opposed to less as we have found. Thus it is possible the synaptic depression induced bursting is more robust to heterogeneity than adaptation induced bursting. However, in this study the heterogeneity was via a uniform distribution in the applied current (as opposed to the Gaussian distributions we consider) and typically $\meanIapp$ was close to rheobase, which could also be factors in their results.

In addition to area CA3 in the hippocampus, adaptation induced bursting has also been suggested as a possible mechanism for the generation of velocity controlled oscillators (VCO's) in the entorhinal cortex by \cite{Hasselmo}.  The VCO's burst at frequencies that vary with the velocity of the animal.  When a subset of VCO's signals are linearly added to a readout neuron, an interference pattern emerges and a grid cell is formed.   \cite{Hasselmo} use a recursively coupled network of homogeneous Izhikevich neurons with adaptation variables given by $W_{jump} = 100$, and $1/\tau_W = 0.03$, and these parameters are strong enough so that adaptation induced bursting can occur.   The network acts a single velocity controlled oscillator, and the burst frequency is set to vary with the velocity of an animal.  This is done by fixing the $g_{syn}$ parameter at a specific value, and inverting the $F(I)$ curve where $F$ is the frequency of bursts, and $I$ is the homogenous applied current to each neuron.   Grid cells can be generated by using multiple networks and linearly adding their output currents to a read-out neuron.   This was done under uncorrelated noisy inputs arriving to each neuron.   However, \cite{Hasselmo} state that synchrony in the noise (which can come from the animals velocity signal for example) coming to each VCO network can disrupt grid-cell formation.  Here, we have shown that a heterogeneous network of oscillators can still maintain a network level oscillation rate, even if the individual neurons have different behaviors.     The network level behaviors are predicted from the mean-field systems.   Given the fact that the individual neurons are heterogeneous, any synchronized noise input into the individual neurons should become increasingly desynchronzied by the differing resposnes of the individual neurons.   As a network level oscillation exists and the heterogeneity will likely desynchronize any noise coming to the individual oscillators, this is a plausible means of generating velocity controlled oscillators.  We leave this particular application of mean-field theory for future work.  

In either of these applications, a mean-field system may yield valuable insights as to the mechanisms of bursting and the parameter regions they occur in.  By carefully choosing the appropriate bifurcation parameters, and accounting for the level of heterogeneity in the neurons in the network, one can determine the bifurcation types and behaviors neurons in these different networks display, in addition to estimates of the different distributions to yield insights about the real cells.

\appendix
\section*{Appendix A: Computing $\langle v\rangle$ and $\langle v | \bm\beta \rangle$}\label{app0}

When the quasi-steady state approximation is applicable, a very convenient method emerges for computing the moments of $v$.  In particular, the quasi-steady state approximation not only yields the steady state flux, but it also yields the steady state density: 
\begin{eqnarray*}
\rho(v) =  \frac{J(s,\langle w \rangle)}{F(v) - \langle w \rangle + gs(e_r-v) + I} 
\end{eqnarray*}
where $J(s,\langle w \rangle)$ is the steady state flux ($\langle R_i(t)\rangle$ for $H(\langle w \rangle,s)\geq 0$).   Obviously with the density, we can compute the moments of any arbitrary function of $v$.  For example, $\langle v \rangle$ is given by 
\begin{eqnarray}
\langle v \rangle =  \int_V\frac{J(s,\langle w \rangle)v\,dv}{F(v) - \langle w \rangle + gs(e_r-v) + I} \quad \text{if} \quad H(s,\langle w \rangle)\geq 0\label{mevo1}
\end{eqnarray}
However, once we cross the switching manifold, equation (\ref{mevo1}) is no longer valid.  An approximation has to be made to compute $\langle v\rangle$ in this region.  In particular, we will assume that $\langle F(v)\rangle = F(\langle v \rangle)$, and that the dynamics of $\langle v\rangle$ are fast relative to $s$ and $\langle w \rangle$.  If so, then the dynamics of $\langle v \rangle$ are approximately given by 
\begin{equation}
\langle v \rangle' \approx F(\langle v \rangle) -\langle w \rangle + gs(e_r - \langle v \rangle) + I 
\end{equation}
and we can solve the steady state equation for $\langle v \rangle$
$$F(\langle v \rangle) -\langle w \rangle + gs(e_r - \langle v \rangle) + I = 0$$ 
However, we have to be careful here. Based on the assumptions on $F(v)$ we know that there are only two solutions to this system, and we need to solve for the stable one.  For the Izhikevich neuron for example, this is given by 
\begin{equation}
\langle v_-\rangle  = \frac{\alpha+gs}{2} - \sqrt{-H(s,\langle w \rangle)}\label{mev2}
\end{equation}
Computation of the integral \label{mev1} for the Izhikevich neuron, in conjunction with equation \ref{mev2} yields the following equation for $\langle v \rangle$ over the entire $s,\langle w\rangle$ plane: 
\begin{equation}
\langle v \rangle =  \left\{  \begin{array}{lr}  \frac{\langle R_i(t)\rangle}{2}\log\left(\frac{(v_{peak} - \frac{\alpha+gs}{2})^2 +H(s,w)}{(v_{reset} - \frac{\alpha+gs}{2})^2 +H(s,w)}\right) + \frac{\alpha+gs}{2}   & : H(\langle w \rangle,s)\geq 0  \\      \frac{\alpha+gs}{2} - \sqrt{-H(s,w)}  &:H(\langle w \rangle,s)<0   \end{array}   \right.  \label{mev3}
 \end{equation}
The same formulas can actually be used for $\langle v |\bm\beta\rangle$, with the interpretation that they are now explicit functions of $\bm\beta$  In particular, one can show that under tonic firing, $\langle v | \bm\beta\rangle$ is given by the same formula as $\langle v \rangle$ when $H(\langle w \rangle,s,\bm\beta)\geq 0$.  Additionally, the formula for $\langle v \rangle$ for $H(s,\langle w \rangle,\bm\beta)<0$ is also a satisfactory approximation for $\langle v|\bm\beta\rangle$.  To once again reiterate that while the formula for the mean $\langle v|\bm\beta\rangle$ under heterogeneity is the same as the formula for $\langle v\rangle$ under homogeneity, this does not imply that $\langle v \rangle = \langle v|\bm\beta\rangle$.  In fact, to compute $\langle v\rangle$, one has to compute the traditional integral for it:
$$ \langle v\rangle =\int_\beta \langle v|\bm\beta \rangle \rho_\beta(\bm\beta)\,d\bm\beta $$ 

\section*{Appendix B: Computing the Proportion of Bursting Neurons in Direct Simulations}\label{appA}

Performing direct simulations of a network is fairly straight forward, but it is 
somewhat difficult to use the results of the simulations to automatically 
classify a given neuron, let alone the entire network, as bursting or 
tonically firing. We considered various classifiers and opted to use the ratio of 
the largest to the smallest interspike interval for a neuron when the 
network has reached steady state.   For a bursting neuron, the ratio of its 
largest interspike interval (which is the interburst interval) to its smallest 
interspike interval should be large.  Thus for the $i^{th}$ neuron we define 
\begin{equation}
\lambda_i = \frac{\max ISI}{\min ISI} 
\end{equation}
where the max/min is determined after a suitable time period of steady state behaviour (either bursting or tonically firing).  We set a critical ratio, $\lambda_c$, and classify neuron $i$ as bursting if $\lambda_i$ is higher than this ratio.  The critical ratio is typically taken to be 2.  This is the single neuron classifier.  We use the single neuron classifier to make an estimate for the total proportion of bursting neurons in the network as follows
\begin{equation}
p_{burst} = \frac{1}{N}\sum_{i=1} H(\lambda_i - \lambda_c)\label{pbursteq}
\end{equation}
where $H$ is the Heaviside function.    We can use $p_{burst}$ to classify a network by picking some specific critical value of $p_{burst}$ as a threshold for a classifier, however it is more informative to simply plot the contours of $p_{burst}$ for a set of simulations run over a mesh in the bifurcation parameters of interest.

\section*{Appendix C: Multiple Subpopulations via Mixed Distributions of Heterogeneity}\label{appB}

In \cite{us2} we considered multiple subpopulations for heterogeneous networks by simulating discrete homogeneous subpopulations within a network.   For example, we simulated two subpopulations with two different sets of parameters corresponding to weakly adapting and strongly adapting neurons as a first attempt at studying inhomogeneous networks.  However, a more realistic way of analyzing subpopulations in heterogeneous networks, is through the use of mixed distributions. 
A mixed random variable $Z$ is a function of two or more random variables. 
For example, if $X$ and $Y$ are random variables with probability density 
functions $f_X$ and $f_Y$, respectively, then 
\begin{equation}
Z = \begin{cases} X & \text{with probability $p$} 
   \\ Y & \text{with probability $1-p$} 
\end{cases}
\end{equation} 
is a mixed random variable with probability density function 
\begin{equation}
f_Z(z) = pf_X(z) + (1-p)f_Y(z). \label{mixedpdf}
\end{equation}

Now consider a network where the adaptation jump size $w_{jump}$ comes from two different subpopulations.  In one subpopulation, $\langle w_{jump}\rangle$ is large, and in the other, $\langle w_{jump}\rangle$ is small.   We can simulate these two subpopulations within an individual (all-to-all) coupled network using heterogeneity, and the density function 
\begin{equation}
f_{wjump}(w) = p_{SA}f_{SA}(w) + (1-p_{SA})f_{WA}(w)
\end{equation}
where $p_{SA}$ denotes the proportion of strongly adapting neurons and $SA$ denotes the strongly adapting subpopulation, and $WA$ denotes the weakly adapting subpopulation.  Note that these densities have higher moments than simply $\langle w_{jump}\rangle$, and it is the magnitude of these moments that determine whether or not the density in the heterogeneous parameter we are considering is bimodal, indicative of multiple subpopulations.   Using this approach we can analyze how the bimodality affects the steady state distribution properties.   We can also use more of the data gained from real networks for the purposes of simulation, as opposed to arbitrarily classifying neurons as strongly adapting or weakly adapting.  The parameters of the individual density functions can be approximated, along with $p$, using standard statistical approaches (maximum likelihood estimation, for example).   More importantly however, the same mean-field equations apply to a unimodal or a multimodal distribution of heterogeneity.

\section*{Appendix D:  Differentiation and Numerical Inversion of Parameter Distributions}\label{appC}

The procedure in section~\ref{mf3sec} requires not only the estimation of 
$\meanRb$, but also the calculation of its inverse and the derivative of this
inverse. To calculate $\meanRb$ the mean field equations 
\eqref{wone3}-\eqref{MF3R} are discretized in $\beta$ and the resulting 
equations are numerically 
integrated to steady state, which yields the steady state value of $\langle R_i | \beta_j\rangle$ at each mesh point, $\beta_j$. The inverse $g^{-1}(r)$ is 
calculated via numerically inverting the steady state $\langle R_i| \beta_j\rangle$ as a function of $\beta_j$ as follows. The MATLAB function interp1 is used to interpolate values of $\beta$ given values of $R$ using the steady state $(\langle R_i |\beta_j \rangle,\beta_j)$ mesh points.   The derivative of the inverse is calculated using a 
finite-difference approximation over the mesh:
$$ \frac{d g^{-1}}{dr}\bigg|_{r = r_j} 
\approx \frac{g^{-1}(r_{j+1}) - g^{-1}(r_{j})}{r_{j+1}-r_j}$$ 
This is then used to find $\rho_R(r)$ at each mesh point via equation
\eqref{invmagic}.

To implement the computations in section~\ref{invsec}, equations 
\eqref{woneI}-\eqref{MF3I} are discretized 
in $\beta$ and numerically integrating to compute the steady state value of 
$\meanRb$ at each mesh point, $\langle R_i|\beta_j\rangle$. 
This is then used to compute $\frac{d}{d\beta}\langle R | \bm\beta \rangle$ 
through a first order finite-difference approximation over the discrete mesh: 
$$ \frac{d\langle R_i | \bm\beta \rangle}{d\bm\beta}\bigg|_{\beta = \beta_j} \approx \frac{ \langle R_i | \bm\beta_{j+1}\rangle -  \langle R_i | \bm\beta_{j}\rangle}{\beta_{j+1}-\beta_j}.$$ 
These two quantities are then used as approximations of $g(\beta)$ and 
$g'(\beta)$ in equation \eqref{magic} to find $\rho_{beta}(\beta)$ at each mesh point.

To estimate the parameter values for individual neurons, we take the discretized
steady state firing rate, $\langle R_i | \bm\beta_j\rangle$, calculated as 
indicated above.  We then invert the functional relationship between 
$\langle R_i | \bm\beta_j\rangle$ and $\beta_j$, and interpolate the 
$\beta$ values of the individual neurons using their firing rate, i.e., we 
treat $\langle R_i | \bm\beta_j \rangle$ and $\beta_j$ as the $(x_j,y_j)$ points 
to be interpolated.  
This yields an estimate of the parameter values for each individual neuron, unlike in the approach for section~\ref{mf3sec} which yielded
an estimate of the overall distribution.  Note that all that is required for this computation is knowledge of the steady state firing rate for each neuron. 
We use the griddata and the interp1 functions in MATLAB to perform the interpolation.

\renewcommand{\abstractname}{Acknowledgements}
\begin{abstract}
This work benefitted from the support of the Natural Sciences and Engineering Research Council of Canada and the Ontario Graduate Scholarship program
\end{abstract}


\end{document}